\documentclass{article}

\usepackage{arxiv}

\usepackage[utf8]{inputenc} % allow utf-8 input
\usepackage[T1]{fontenc}    % use 8-bit T1 fonts
\usepackage{booktabs}       % professional-quality tables
\usepackage{amsfonts}       % blackboard math symbols
\usepackage{nicefrac}       % compact symbols for 1/2, etc.
\usepackage{microtype}      % microtypography
\usepackage{lipsum}		% Can be removed after putting your text content
\usepackage{natbib}
\usepackage{doi}

\usepackage{graphicx}
\usepackage{float}
\usepackage{subfigure}
\usepackage{titlesec}
\usepackage[ruled,vlined]{algorithm2e}
\usepackage{array}
\usepackage{mdwmath}
\usepackage{hyperref}
\usepackage{url}
\usepackage{ulem}

\usepackage{amssymb}
\usepackage{amsmath}
\usepackage{multicol}
\usepackage[figuresright]{rotating}
\usepackage{bm}
\usepackage{caption}
\usepackage{fancyhdr}

\title{
 XMAM:X-raying Models with A Matrix to Reveal Backdoor Attacks for Federated Learning
}

\author{ \href{https://orcid.org/0000-0001-8765-053X}{\includegraphics[scale=0.06]{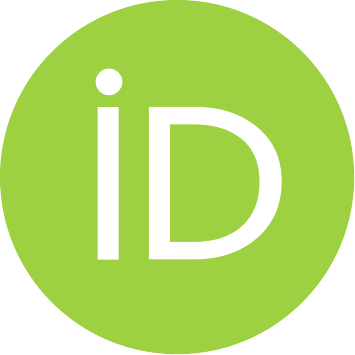}\hspace{1mm}Jianyi Zhang}\thanks{Corresponding author: \texttt{zjy@besti.edu.cn, xiali.hei@louisiana.edu} } \\
	Beijing Electronic Science and Technology Institute\\
	Beijing, China 100070 \\
	\And
	{\hspace{1mm}Fangjiao Zhang, Qichao Jin, Zhiqiang Wang} \\
	Beijing Electronic Science and Technology Institute\\
	Beijing, China 100070 \\
	\And
	{\hspace{1mm}Xiaodong Lin} \\
	University of Guelph\\
	Ontario N1G 2W1, Canada \\
 \And
	{\hspace{1mm}Xiali Hei*} \\
	University of Louisiana at Lafayette\\
	Louisiana US 70503 \\
}

\renewcommand{\shorttitle}{XMAM}

\hypersetup{
pdftitle={XMAM:X-raying Models with A Matrix to Reveal Backdoor Attacks for Federated Learning},
pdfsubject={q-bio.NC, q-bio.QM},
pdfauthor={Jianyi Zhang, Fangjiao Zhang, Qichao Jin, Zhiqiang Wang, Xiaodong Lin, Xiali Hei},
pdfkeywords={First keyword, Second keyword, More},
}

\begin{document}
\maketitle

\begin{abstract}

Federated Learning (FL), a burgeoning technology, has received increasing attention due to its privacy protection capability. However, the base algorithm FedAvg is vulnerable when it suffers from so-called backdoor attacks. Former researchers proposed several robust aggregation methods. Unfortunately, due to the hidden characteristic of backdoor attacks, many of these aggregation methods are unable to defend against backdoor attacks. What's more, the attackers recently have proposed some hiding methods that further improve backdoor attacks' stealthiness, making all the existing robust aggregation methods fail. 

To tackle the threat of backdoor attacks, we propose a new aggregation method, \underline{X}-raying \underline{M}odels with \underline{A} \underline{M}atrix (XMAM), to reveal the malicious local model updates submitted by the backdoor attackers. Since we observe that the output of the $Softmax$ layer exhibits distinguishable patterns between malicious and benign updates, unlike the existing aggregation algorithms, we focus on the $Softmax$ layer's output in which the backdoor attackers are difficult to hide their malicious behavior. Specifically, like medical X-ray examinations, we investigate the collected local model updates by using a matrix as an input to get their $Softmax$ layer's outputs. Then, we preclude updates whose outputs are abnormal by clustering. Without any training dataset in the server, the extensive evaluations show that our XMAM can effectively distinguish malicious local model updates from benign ones. For instance, when other methods fail to defend against the backdoor attacks at no more than 20\% malicious clients, our method can tolerate 45\% malicious clients in the black-box mode and about 30\% in Projected Gradient Descent (PGD) mode. Besides, under adaptive attacks, the results demonstrate that XMAM can still complete the global model training task even when there are 40\% malicious clients. Finally, we analyze our method's screening complexity and compare the real screening time with other methods. The results show that XMAM is about 10-10000 times faster than the existing methods.

\end{abstract}


\section{Introduction}

Federated Learning (FL) \cite{konevcny2016federated,mcmahan2017communication} is a burgeoning technology. To sum up, FL possesses the following three characteristics, which are distinguished from standard distributed learning: (i) The server gathers clients' local models instead of their training data to train a global model jointly. (ii) The distribution of client training data is non-i.i.d. (iii) The server cannot control the training processes of clients. With the promulgation of the privacy regulations General Data Protection Regulation (GDPR), FL has received increasing attention and applications due to its privacy protection capability. 

The first FL aggregation method is FedAvg \cite{mcmahan2017communication} proposed by Google. However, the base algorithm FedAvg is vulnerable when there malicious clients exist in the training process. In FL, a malicious client inducing the global model to misclassify the data selected by the attacker but guaranteeing the convergence of the global model is called backdoor attack \cite{bagdasaryan2020backdoor} (\textit{i.e.}, targeted attack \cite{kairouz2019advances}). A malicious client can use data poisoning attacks, which poison the client's local training data, or local model poisoning attacks, which modify the client's local model update, to achieve his/her purpose. In this paper, we focus on the backdoor attacks (Trigger attack \cite{gu2017badnets}, Semantic attack \cite{bagdasaryan2020backdoor}, and Edge-case attack \cite{wang2020attack}) since they are more challenging problems in FL. We also give some explanations of our method's performance on the adaptive attacks. 

There are still many deficiencies in existing aggregation methods. To make the malicious local model updates more concealing, attackers proposed some hiding techniques, making them indiscernible from the benign local model updates. The former researchers proposed several existing robust aggregation methods. All of them focus on the local model updates' parameters. For instance, Krum\cite{blanchard2017machine} computes the Euclidean distances between local model updates and selects the one with the smallest distance as the global update. However, the aggregation methods based on Euclidean distance will be difficult to distinguish the malicious local model updates from the benign ones when the backdoor attacks are under hiding modes (\textit{e.g.}, Projected Gradient Descent (PGD) mode \cite{wang2020attack} and Stealthy Model Poisoning (SMP) mode \cite{bhagoji2019analyzing}). FLTrust \cite{cao2020fltrust} performs well in defending against a more significant proportion of malicious clients. Still, it requires the server to possess a small batch of training data, which might be impractical in some fields (\textit{e.g.}, financial and digital health\cite{rieke2020future}) because the private local training data would be challenging to obtain or fabricate. 

\textbf{Our work:} To tackle the threat of backdoor attacks that use hiding techniques, we proposed a new aggregation method, \underline{X}-raying \underline{M}odels with \underline{A} \underline{M}atrix (XMAM). Like an X-ray in medical examinations, we use a matrix to examine the local model updates. Specifically, we directly utilize the submitted local model updates to be the parameters of the network and input a matrix (a random matrix is enough) to get the $Softmax$ layer's outputs. Then, we preclude local mode updates whose $Softmax$ layer's outputs are abnormal by clustering. In our method, the malicious local model updates generated by backdoor attacks that use hiding techniques can be easily distinguished from the benign local model updates. 

\textbf{XMAM can defend against existing attacks:} We evaluate our aggregation method's superiority on three backdoor attacks (Trigger attack \cite{gu2017badnets}, Semantic attack \cite{bagdasaryan2020backdoor}, and Edge-case attack \cite{wang2020attack}) under Black-box mode \cite{wang2020attack}, PGD mode \cite{wang2020attack}, and SMP mode \cite{bhagoji2019analyzing}. Note that the Black-box mode is the initial mode that the attackers do not use hiding techniques. The PGD and SMP modes are two advanced modes that the attackers use hiding techniques. Compared with six existing aggregation methods (FedAvg, NDC, RSA, RFA, Krum, and Multi-Krum), we can find that some methods fail even when the backdoor attacks are in the Black-box mode, and other methods fail when the backdoor attacks are in PGD or SMP mode. Only Krum and our method successfully defend against backdoor attacks, whatever the mode is. Our method performs best since Krum only collects one local update as the global update per iteration. Although this allows Krum to avoid malicious local model updates, it also makes global model convergence very slow and reduces accuracy. Furthermore, we evaluate our method on two adaptive attacks: the Krum attack and the XMAM attack. The first one is proposed by \cite{fang2020local}, and the latter one is designed by ourselves according to the framework of the Krum attack. The results manifest that our method is resilient to adaptive attacks. 

Our contributions can be summarized as follows:
\begin{itemize}
\item To the best of our knowledge, our work is the first to reveal backdoor attacks in FL by focusing on the $Softmax$ layer's outputs instead of the local model updates. The experimental results demonstrate that the $Softmax$ layer's output can reflect a model's information, and different models show different information.
\item We successfully tackle the threat of backdoor attacks that use hiding techniques. The experimental results manifest that our method is more effective in detecting malicious local model updates than other existing robust aggregation methods.
\item Our method is not vulnerable when the attackers implement adaptive attacks, which further demonstrates the robustness of our method.
\item Our method is time-saving in the process of detection compared with other methods since we mainly focus on an $M$-dimensional space but others focus on a $\zeta$-dimensional space, where $M$ is the number of classes of data and $\zeta$ is the number of parameters of local model update.
\end{itemize}

\section{Background}
\subsection{Federated Learning (FL)}

In FL, multiple clients jointly train a global model by their imbalanced local dataset. Ideally, the optimization model is as follows:

\begin{equation}
\min _{{w}}\left\{F({w}) = \sum_{i=1}^{N} \frac{1}{N} F_{i}({w})\right\}
\end{equation}

where $N$ is the number of clients. $F_i(\cdot)$ is the local objective, which is defined by  

\begin{equation}
F_{i}({w}) = \frac{1}{|\mathcal{D}_{i}|} \sum_{j=1}^{|\mathcal{D}_{i}|} \mathcal{L}\left({w} , x_{i, j}\right)
\end{equation}

where $\mathcal{L}(\cdot,\cdot)$ is a user-specified loss function, and the $i^{th}$ client holds the $|\mathcal{D}_{i}|$ training data: $x_{i,1}, x_{i,2}, \ldots, x_{i,|\mathcal{D}_{i}|}$. 

\begin{figure}[ht]
\begin{center}
   \includegraphics[width=0.7\linewidth]{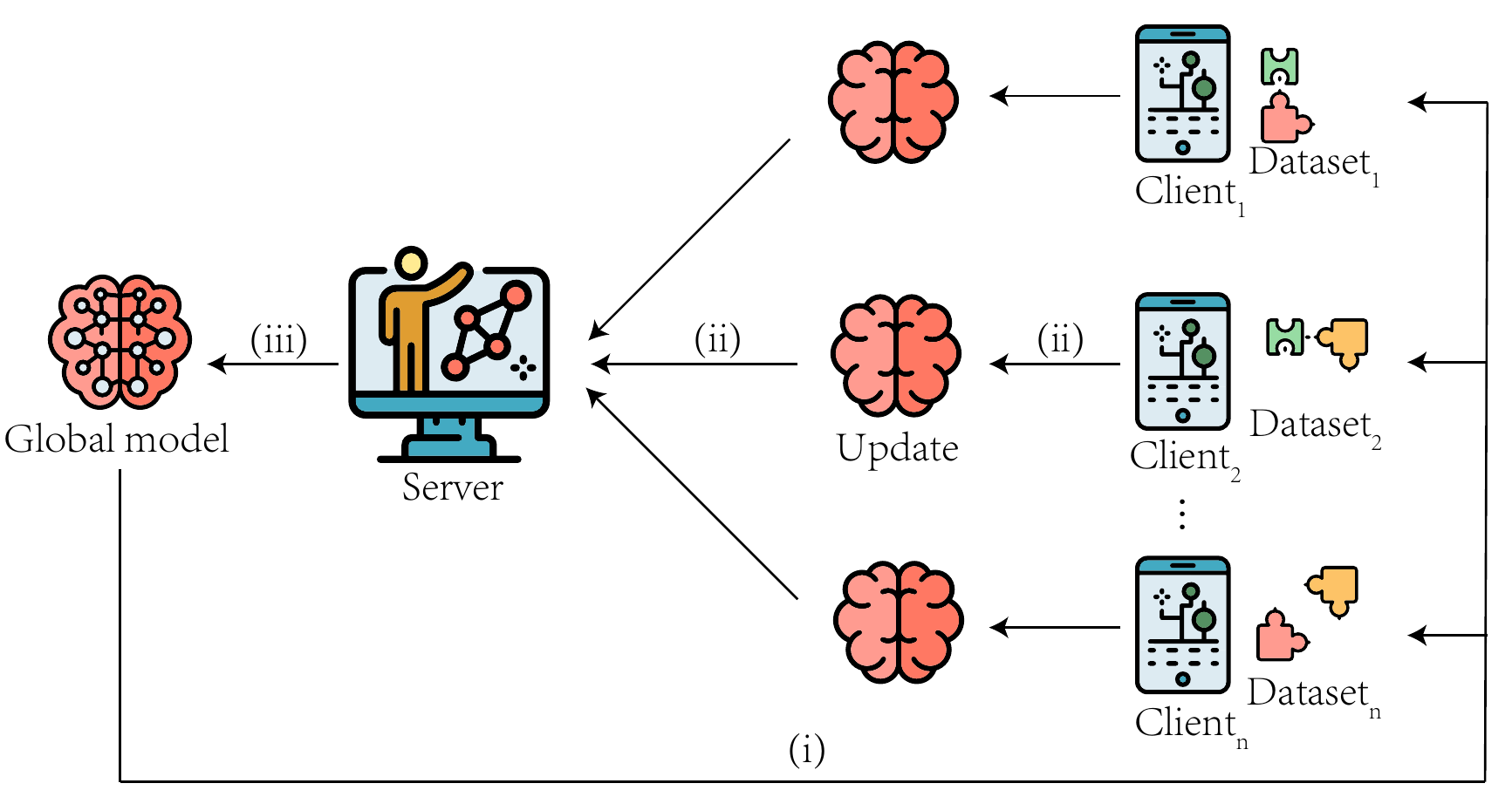}
\end{center}
   \caption{Illustration of the three steps in one iteration of FL. There are $N$ clients (\textit{e.g.}, smartphones or edge devices) and a server (service provider, \textit{e.g.}, Google or Amazon). Each client has different classes and number of data, representing the heterogeneous distribution of client data.}\vspace{-4mm}
\label{fig:FL_conception}
\end{figure}

We roughly divide FL into three steps in one iteration (illustrated in Fig. \ref{fig:FL_conception}): (i) The server sends the aggregated global model to clients. (ii) The clients update the global model by local training dataset and return the local model updates to the server. (iii) The server collects a portion of clients' local model updates and aggregates them to update the global model for the next iteration. When the server is the only defender, the only way to resist poisoning attacks is by focusing on step (iii). So the following definitions of all methods account for step (iii). Note that the global model and the local model update we said in this paper are parameters.

\subsection{Backdoor attacks in federated learning}

Generally, obtaining a high accuracy on the testing dataset is a model's main task, and the main task can be considered as a series of sub-tasks, like performance on a certain type of data. Manipulating the performance of a model on special types of data is the so-called backdoor attack. Backdoor attacks (targeted attacks \cite{kairouz2019advances}) do not destroy the global model's accuracy on the main task but induce it to make attacker-chosen mistakes on backdoor tasks. In the paradigm of FL, the server has no power to inspect the cleanliness of clients' data. Thus, a malicious client might poison his local data and train a malicious local model update based on it. When the server frequently receives the malicious local model updates, the global model will compromise on backdoor tasks once received from malicious clients and aggregate them.

\begin{figure}[ht]
\begin{center}
\includegraphics[width=0.6\linewidth]{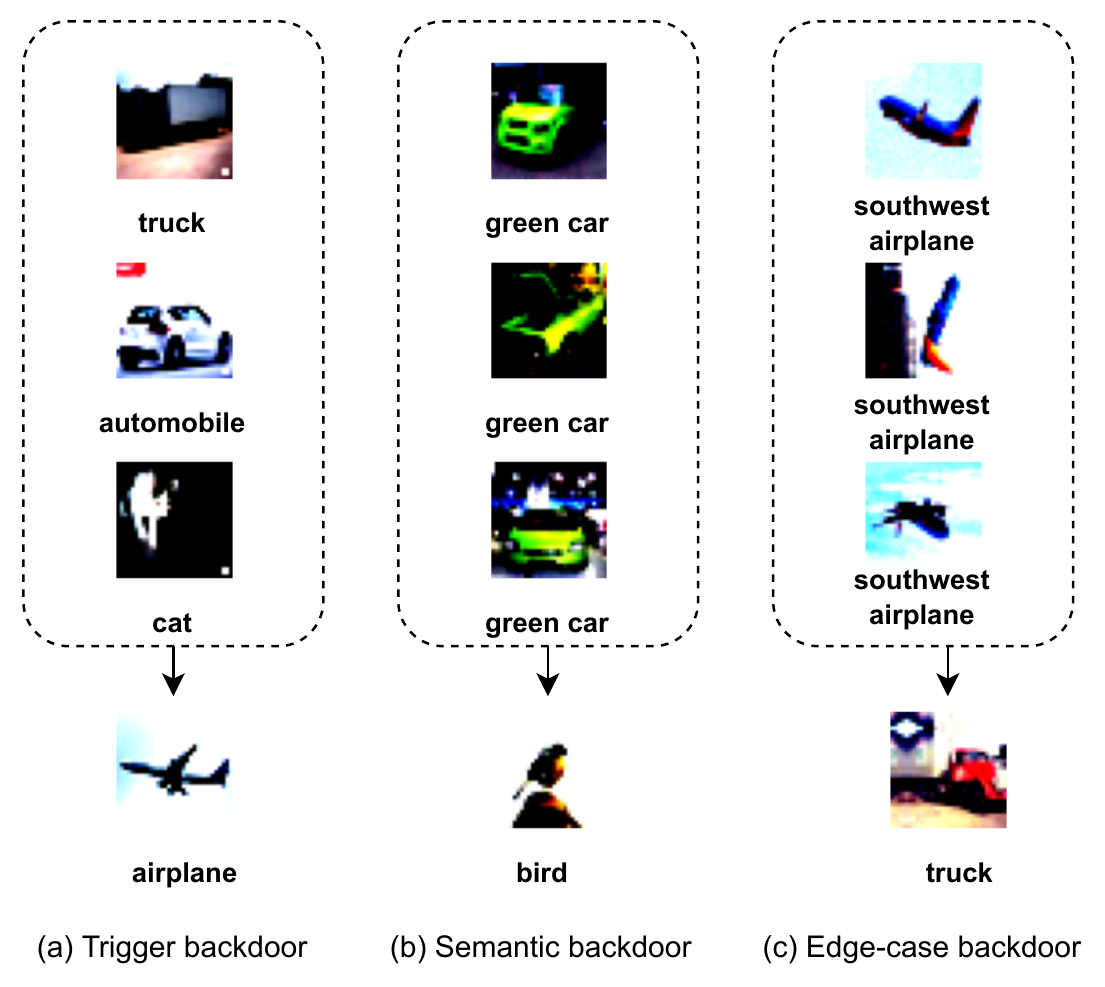}
\end{center}

\caption{Illustration of the three backdoor attacks in FL. For each backdoor attack, we give an example. For the Trigger attack, we stamp a white pixel block on the corner of the images and label them as airplanes. For the Semantic attack, we use the images of green cars and label them as birds. For the Edge-case attack, we use the images of southwest airplanes, which are rare data, and label them as trucks.}
\label{fig:backdoor_example}
\end{figure}

Currently, there are three typical backdoor threats in FL: Trigger backdoor, Semantic backdoor, and Edge-case backdoor. Trigger backdoor assumes the malicious clients poison their local data by stamping a pattern on the images and modifying the labels to any class they want. Then, the ultimate global model will make an attacker-chosen judgment on data that has the same pattern on it. Therefore, the Trigger backdoor requires a data modification both on the local model's training period and the global model's inference period. The Semantic backdoor does not require a data modification on the global model's inference period. It directly exploits the special feature (\textit{e.g.}, a green car, a car painted with stripes,  and a car with stripe background on it) on some data and modifies their label to any class the attacker wants. Then, the ultimate global model will excessively learn the special feature and make a prejudicial judgment on data that contains it. The latest backdoor is the Edge-case backdoor, which directly uses data rarely seen in clients' datasets and modifies their labels to make the ultimate global model to misclassify those data. Edge-case backdoor further reveals the security issue of FL on rare data. Fig. \ref{fig:backdoor_example} illustrates the concrete operations of three backdoor attacks.

\textbf{Why do backdoor attacks threaten FL so much?} 
Earlier, the attackers want to embed the backdoor to the global model in one shot by only one malicious client. Therefore, they amplify the malicious local model update optimized by the poisoned local data to mitigate the effectiveness of other benign clients' local model updates. 
In other words, the global model will be replaced by the malicious client's model in just one aggregation, which is the so-called model replacement technique \cite{bagdasaryan2020backdoor}. Assuming a malicious local model update trained by the poisoned local dataset is $\hat{u}$, and the server collects $\tau$ local model updates in each iteration. A malicious client who uses the model replacement technique will magnify his $\hat{u}$ before submitting it. It's usually magnified $\tau$ times. However, this naive operation magnifies the anomaly degree of malicious local model update so that most current robust aggregation methods can detect and preclude it. For example, an aggregation method \cite{blanchard2017machine} can preclude abnormal local model updates based on Euclidean distance.  

\begin{figure}[ht]
\begin{center}
\includegraphics[width=0.5\linewidth]{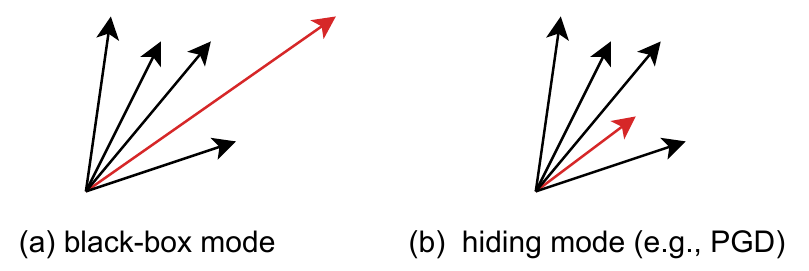}
\end{center}
\caption{Illustration of the malicious local model update before hiding and after hiding. The black vectors denote the benign local model updates, and the red vector denotes the malicious local model update. When the malicious client uses hiding techniques, the malicious local model updates scales in magnitude, making it hard to distinguish from benign local model updates in the Euclidean space.} 
\label{fig:vector}
\end{figure}

Then, the attackers find that a group of collusive malicious clients who can jointly attack the global model for several iterations can also successfully embed the backdoor without using model replacement. Since it only needs the malicious clients injecting poisoned data into their clean local dataset without any other manipulations, this attack mode is called the black-box mode \cite{wang2020attack}. Although the malicious local model updates optimized under the black-box mode are more similar to benign local model updates, they might still possess an unusual angle or magnitude compared with the benign ones (Fig. \ref{fig:vector}(a)). Thus, some existing aggregation methods, like Krum and Multi-Krum, can detect it when the number of malicious clients does not surpass 50\% in each iteration. 

To make the malicious local model updates more concealing (Fig. \ref{fig:vector}(b)), the attackers proposed two hiding techniques: PGD model poisoning \cite{wang2020attack} (we call it PGD mode below) and stealthy model poisoning \cite{bhagoji2019analyzing} (we call it SMP mode below). Attacks in PGD mode scale down the malicious local model updates to a smaller norm, which makes them indiscernible from the benign local model updates:

\begin{equation}
\hat{u}^{'} = \frac{\epsilon \hat{u}}{ \left\|  \hat{u}  \right\|_2}
\end{equation}

where $\epsilon$ is the scaling down magnitude and $\hat{u}^{'}$ is the malicious local model update after scaling down. Attacks under SMP mode use a new objective function that considers three losses of a model: the loss on clean data, the loss on poisoned data, and the loss of distance between the model after training and the model before training. With the above three constraints, a model trained in SMP mode is similar to the benign models and maintains its backdoor alive. Therefore, the adversarial objective becomes:

\begin{equation}
\begin{split}
\underset{\hat{w}}{\operatorname{argmin}}\quad   {\rho}_1 L(\mathcal{D}_{p}, & \hat{w})+L\left(\mathcal{D}_{c}, \hat{w}\right)+ {\rho}_2 \left\|\hat{w}-w_{\text {g}}\right\|_{2}
\end{split}
\end{equation}
Then, the malicious local model update $\hat{u}$ in SMP mode is:
\begin{equation}
\hat{u} = \hat{w} - w_{\text{g}}
\end{equation}

where $\mathcal{D}_{p}$ is the poisoned data in the client's local dataset, $\mathcal{D}_{c}$ is the clean data in the client's local dataset, $w_{\text{g}}$ is the global model in the previous iteration, $\hat{w}$ is the malicious client local model after training and ${\rho}_1$ and ${\rho}_2$ are the weights of the objective function.

\subsection{Existing robust aggregation methods}
The existing robust aggregation methods can be roughly divided into three categories: (i) \textit{limiting the update}, (ii) \textit{finding the ``center''}, and (iii) \textit{detecting and precluding}. The typical representatives of the three categories are introduced below. Notations below are listed in Appendix\ref{tab:notations}.

The first category is \textit{limiting the update}. The core of this type of method is to punish the local model update who has a large norm or regularize all the local model updates to a small norm. The representative methods are as follows:

\textbf{NDC\cite{sun2019can}:} \ \ Norm Difference Clipping (NDC) clips the part of local model update whose norm exceeds the threshold $ \frac{\left\| {u} \right\|_{2}}{\delta} $ when it is greater than 1. The clipped local model update is calculated as follows: 
\begin{equation}
{u}_i^c = \sum_{i=1}^{\tau} \frac{{u}_i}{\max(1, \frac{\left\| {u}_i \right\|_{2}}{\delta})}
\end{equation}
where ${u}_i^c$ is the local model update after clipping and $\delta$ is the clipping parameter.

\textbf{RSA\cite{li2019rsa}:} \label{oth:rsa} \ \ To punish the abnormal local model updates, RSA only considers the directions rather than the magnitudes of the local model updates in each iteration. So, all the local models are constricted in a boundary, which can be explained as follows:
\begin{equation}
{u}^{t+1} = \sum_{i=1}^{\tau} {\beta}_{r} Sign( {u}_i^t )
\end{equation}
where $Sign(x)$ equals to $1$ when $x > 0$, $-1$ when $x < 0$, and an arbitrary value within $[-1, 1]$ when $x = 0$. 

The second category is \textit{finding the ``center''}. The core of this type of method is to exploit the local model updates submitted by clients to find a compromised update, which is the ``center'' of the local model updates, to update the global model. The representative methods are as follows:

\textbf{RFA\cite{pillutla2019robust}:} \ \ RFA takes the weighted geometric median of collected local model updates using the smoothed Weiszfeld's algorithm as the aggregated global model. A particular round of the smoothed Weiszfeld's algorithm is computed as follows:
\begin{equation}
q_i^r = \frac{p_i}{v \cup \left\| z^r - {u}_i \right\|_{2}}
\end{equation}
\begin{equation}
z^{r+1} = \frac{ \sum_{i=1}^{\tau} q_i^r {u}_i }{ \sum_{i=1}^{\tau} q_i^r }
\end{equation}
where $p_{i} = \frac{D_i}{|D|}$ and $z^r$ is the geometric median point in $r^{th}$ round. 

The third category is \textit{detecting and precluding}. The core of this type of method is to detect the malicious local model update and preclude it. The representative methods are as follows:

\textbf{Krum\cite{blanchard2017machine}:} \ \ Krum assumes that the server knows the number $f$ of malicious clients in each iteration and then selects the local model update ${u}^{\star}$, which is at the geometric center of $\tau-f-2$ nearest local model updates, as the global model update. The ${u}^{\star}$ is computed as follows:

\begin{equation}
{u}^{\star} = \mathop{\arg\min}_{{u}_i} \sum_{{u}_{j} \in \Omega_{j, \tau-f-2}}\left\|{u}_{i}-{u}_{j}\right\|_{2}^{2}\bigg|_{{u}_{i}=1,2, \ldots ,{u}_{\tau}}
\end{equation}
where $\Omega_{j, \tau-f-2}$ are the set of $\tau-f-2$ local model updates that have the smallest Euclidean distance to ${u}_{j}$.  

\textbf{Multi-Krum\cite{blanchard2017machine}:} \ \ Multi-Krum is a variant of Krum, which collects $\tau-f-2$ clients' local model updates and then integrates them for the global model update.

\section{Problem setup}
\textbf{Threat model:} We have the below assumptions for malicious clients according to \cite{cao2020fltrust,fang2020local}: (i) they have access to the global model of the previous iterations. (ii) They can manipulate their local training data and local model updates in any way. (iii) They can control the local training hyper-parameters such as local learning rate and local training epochs. Furthermore, we have assumptions that (iv) the number of malicious clients is less than 50\% of the total. The operations of (ii) and (iii) also mean that the malicious clients do not know benign clients' local training data and local model updates and can do nothing about the training process of the benign clients. This setting is defined as partial knowledge by\cite{fang2020local}. Moreover, under the condition that the server has no dataset, the (iv) is common in other papers \cite{wang2020attack,blanchard2017machine,sun2019can,li2019rsa,pillutla2019robust}. 

\textbf{Defense goals:} As in \cite{cao2020fltrust}, we evaluate our method from three aspects: \textbf{fidelity}, \textbf{robustness}, and \textbf{efficiency}. For fidelity, we expect our aggregation method does not sacrifice the performance compared with FedAvg when there are no backdoor attacks. For robustness, the goal of our aggregation method is to have comparable performance to FedAvg* (no malicious client participates) under the most powerful backdoor attacks. And for efficiency, we aim to reduce the screening costs to negligible. 

\textbf{Defender’s knowledge and capability:} We assume the server is the only defender and make the following assumptions: (i) The server has no access to the clients' local training data. (ii) The server has full access to the global model and local model updates from all clients in each iteration\cite{cao2020fltrust}. (iii) The server does not know the number of malicious clients\cite{cao2020fltrust,sun2019can,li2019rsa,pillutla2019robust}. (iv) The server has no dataset. Comparing with \cite{fang2020local} that assumes the server has a test dataset to validate the collected models' accuracy, and \cite{cao2020fltrust} that hypothesizes the server has a root dataset (a small dataset which contains about 100 training examples) to train a benchmark update, our assumption (iv) is more practical from the perspective of privacy. 

\section{XMAM overview and design}

\subsection{High-level idea}

\textbf{Motivation:} Although there are multiple robust aggregation methods proposed by former researchers, the backdoor attacks with hiding techniques still pose a threat to FL. A malicious local model update after being scaled down can still embed the backdoor to the global model illustrating that the aggregation methods (\textit{e.g.}, NDC\ cite{sun2019can} and RSA \cite{li2019rsa}) that limit the magnitude of local model updates are infeasible. The methods (\textit{e.g.}, RFA \cite{pillutla2019robust}) that attempt to find a geometric center of the local model updates fail to mitigate the impact of malicious local model updates. The methods (\textit{e.g.}, Krum and Multi-Krum) based on detecting and precluding are effective measures to thoroughly eliminate the effect of malicious local model updates. However, the detection methods of Krum and Multi-Krum lose efficacy when the malicious local model updates become more concealing. To improve the safety of FL, a new aggregation method that can effectively address this problem is desired.  

\begin{figure}[ht]
\begin{center}
\includegraphics[width=0.6\linewidth]{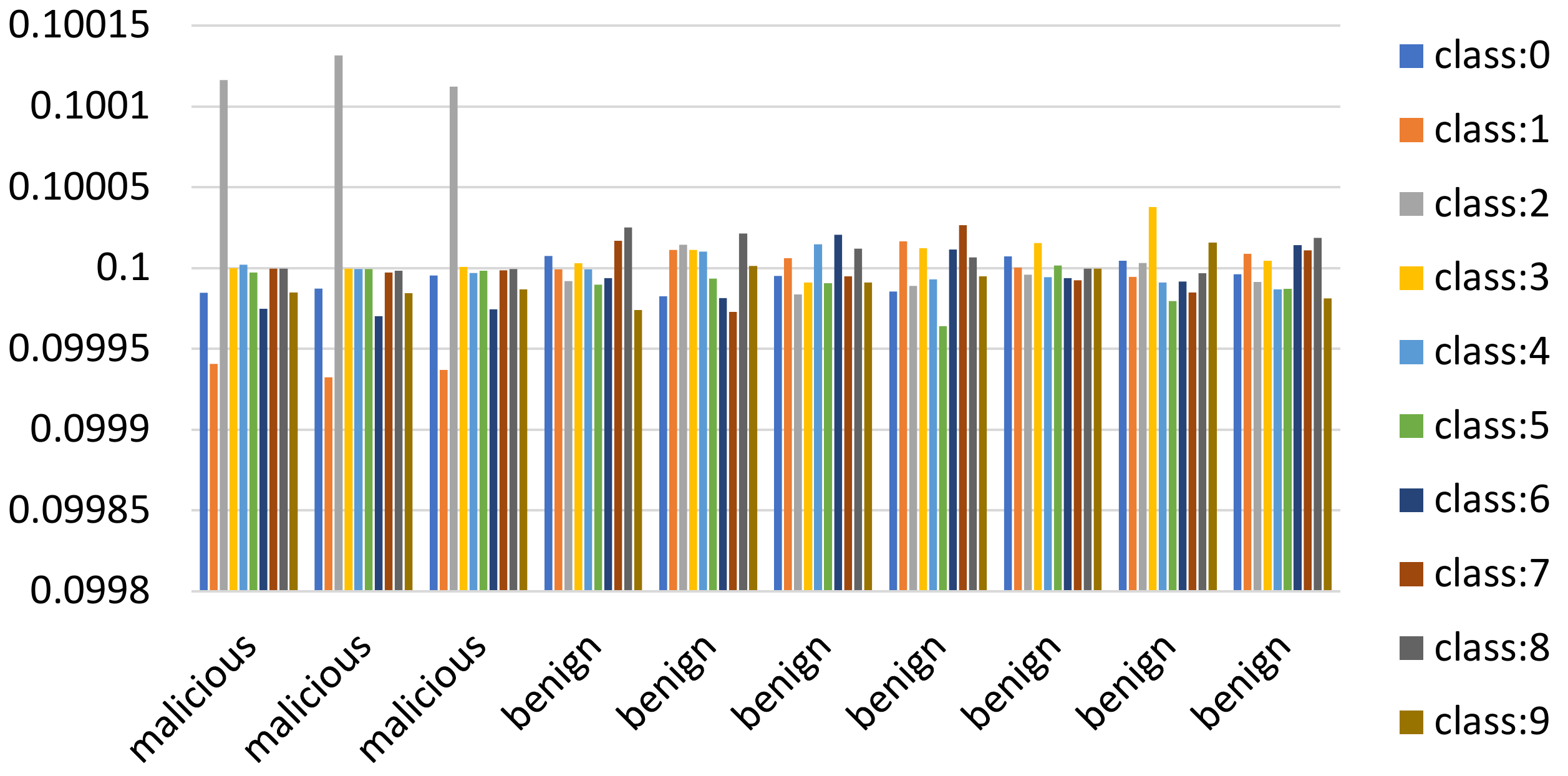}
\end{center}

\caption{SLOUs of ten local model updates. Among them, the first three are SLOUs of malicious clients, and the others are benign clients.}\vspace{-3mm}
\label{fig:intuition}
\end{figure}

\textbf{Challenges:} As mentioned above, we think the most effective way to defend against backdoor attacks is detecting and precluding. Therefore, we follow this idea to design the new method. There are two critical challenges that a new method needs to face. 
\begin{itemize}
\item The first challenge is to distinguish the malicious local model updates from benign ones when the malicious clients are using hiding techniques. 
\item The second challenge is to design the new method without assuming the server has a dataset.
\end{itemize}

%Inspiration and intuition
\textbf{Key observation and idea:} Our key idea is that we directly investigate the collected local model updates by using a matrix as an input to get their $Softmax$ layer's outputs. The elements of this matrix can be random numbers. 

In \ref{app:proof}, we illustrate the feasibility of this method. In the following, we refer to the $\underline{S}oftmax$ \underline{l}ayer's \underline{o}utput of model whose parameter is local model \underline{u}pdate as SLOU.

Our key observation is that the SLOUs exhibits distinguishable patterns between malicious updates and benign ones. For example, Fig. \ref{fig:intuition} shows the SLOUs of updates for ten clients in a certain iteration. Among them, three updates are generated using the Edge-case attack in black-box mode. Furthermore, we plot the dimensionality reduction distribution of local model updates of 100 clients and the dimensionality reduction distributions of 100 corresponding SLOUs in a certain round (Fig. \ref{fig:pgd_eps}). Among them, the local model updates of 20 clients are generated by an Edge-case attack in PGD mode. We found that as the parameter $\epsilon$ of PGD gets smaller and smaller ($\epsilon=1$, $\epsilon=5e^{-1}$, and $\epsilon=5e^{-2}$), that is, the norm of malicious local model updates becomes smaller and smaller, and in the dimensionality reduction distribution of 100 local model updates, 20 malicious local model updates are indistinguishable from benign ones. In this case, the traditional method, such as Multi-Krum, based on Euclidean distance, selects $k$ updates at the center of Euclidean space, and the malicious updates will be selected. In the dimensionality reduction distribution of 100 corresponding SLOUs, there is still a clear distinction between malicious SLOUs and benign SLOUs. Thus, our method can easily preclude malicious updates by clustering.

% Inspired by Knowledge distillation \cite{hinton2015distilling}, in which the teacher network using its $Softmax$ layer's output (soft target probabilities) to teach the student network, we surmise that the $Softmax$ layer's output must contain some critical information of a model. Intuitively, the $Softmax$ layer's output of a benign model and a poisoned model should be different. what's more, the $Softmax$ layer's output is a $M$ dimension probability distribution in which each value is constrained in $[0, 1]$. Therefore, we conjecture that the $Softmax$ layer's output is not severely affected by the norm scaling of a model (i.e., hiding techniques). 

% But, how can we obtain the $Softmax$ layer's output when the server has no dataset? Note that we only need data to obtain the $Softmax$ layer's output instead of expecting a model to learn something. Thus, a random matrix is enough as long as we input the same one. We will mathematically illustrate the effectiveness of our intuition in Appendix \ref{app:proof}.

% We 

% Fig. \ref{fig:intuition} shows the different SLOUs between malicious local model updates and benign local model updates and Fig. \ref{fig:pgd_eps} shows our method's superiority on distinguishing malicious local model updates from benign ones when the malicious local model updates become increasingly small ($\epsilon=1$, $\epsilon=5e-1$, and $\epsilon=5e-2$).

\subsection{XMAM design}

\begin{figure*}[ht]
\begin{center}
   \includegraphics[width=0.8\linewidth]{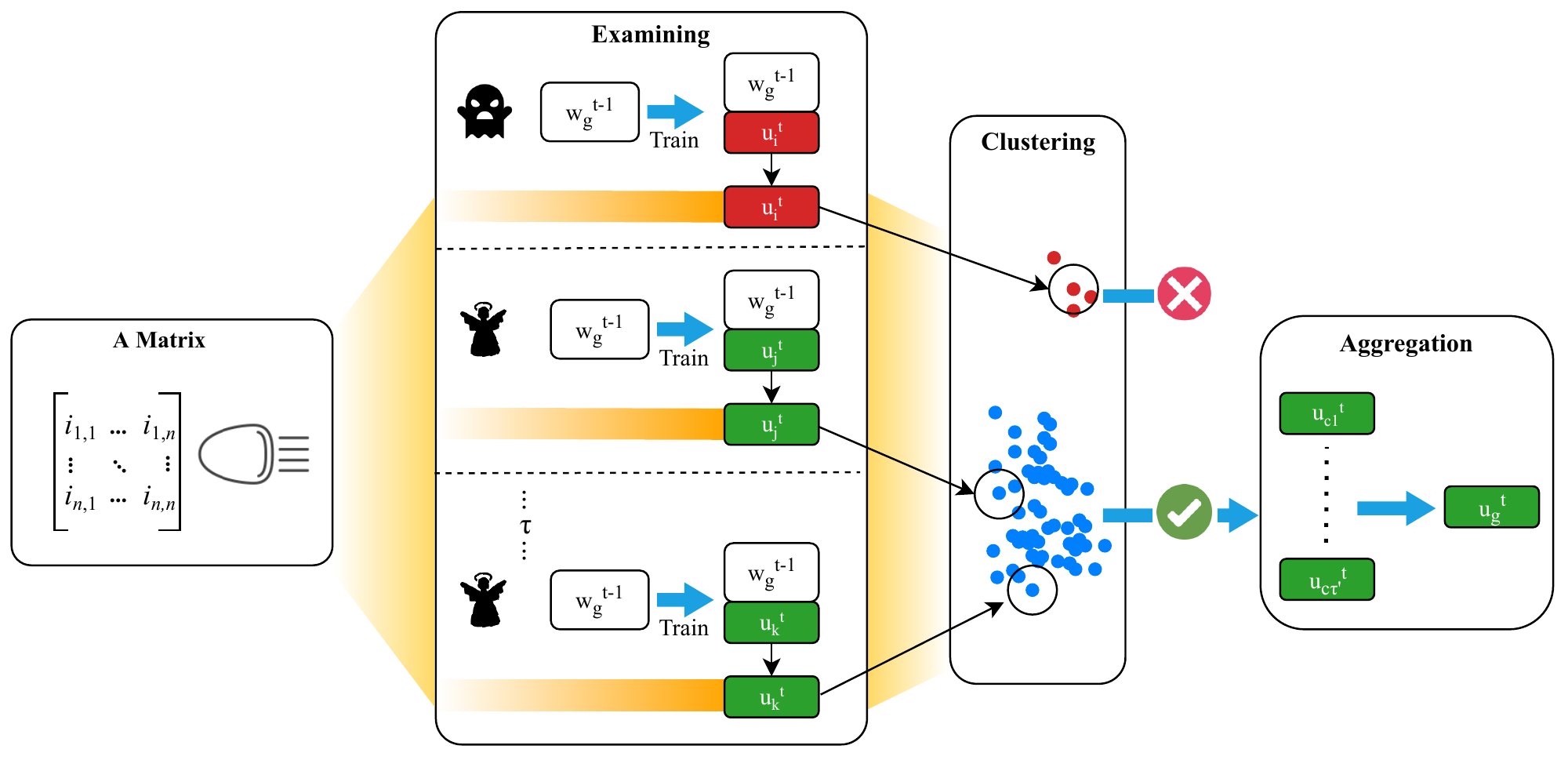}\vspace{-3mm}
\end{center}
   \caption{The procedure of XMAM. After receiving the ${t-1}^{th}$ iteration global model $w_{\text{g}}^{t-1}$ and training based on it, the $\tau$ clients ($f$ of them are malicious clients) submit their local model updates $u_{c1}^{t}, \cdots, u_{c{\tau}}^{t}$ to the server. Note that in the procedure Examining, we only draw three clients $i$, $j$, and $k$ and the three local model updates $u_i^t$, $u_j^t$, and $u_k^t$ submitted by them. After getting the SLOUs using a matrix input, the server clusters these SLOUs and preserves these local model updates ($u_{c1}^{t}, \cdots, u_{c{\tau^{'}}}^{t}$) whose SLOUs are in the major cluster, and aggregates them as the global update $u_{\text{g}}^t$.} \vspace{-4mm}
\label{fig:xmam} 
\end{figure*}

Our new aggregation method consists of three parts: EXAMINING (the server examines the received local model updates using a matrix as input to get the SLOUs), CLUSTERING (the server clusters these SLOUs, and decides which local model updates are benign and should be preserved according to the clustering result), and AGGREGATION (the server aggregates the preserved local model updates and uses them to update the global model). Fig. \ref{fig:xmam} illustrates the process of our aggregation method.

\textbf{EXAMINING:} Considering that the server has no dataset, we generate a matrix and use it as input to examine all the local model updates like an X-ray in medical examinations. Using the same matrix as an input, the server can extract all the local model updates' features (\textit{i.e.}, the SLOU)  in each iteration. The process of EXAMINING is defined as follows:  

\begin{equation}
{SLOU}_i^t = Network({u}^{t}_{i}, \mathcal{D}_{mat})
\end{equation}
where ${SLOU}_i^t$ denotes the $i^{th}$ client local model update's SLOU at $t^{th}$ iteration, $\mathcal{D}_{mat}$ is the matrix, and the function $Network(\cdot, \cdot)$ means inputting data and getting the SLOU.

\textbf{CLUSTERING:} After procedure EXAMINING, the server obtains the SLOUs of the received local model updates. Notice that we have no knowledge of how many local model updates in them are malicious. In other words, there might be no malicious local model updates at all. In addition, the local model updates affected by the non-i.i.d. of clients' local datasets are innately biased. Therefore, we cannot purely divide the local model updates into two clusters. That is to say, we cannot use a clustering algorithm like K-means in which we need to define the number of clusters in advance. Therefore, we use HDBSCAN \cite{campello2013density} as our clustering algorithm since it automatically divides clusters based on node density. After clustering, the server only preserves the local model updates in the major cluster with the maximum number of nodes. The precluded updates might be malicious local model updates or biased benign local model updates. We can express it as follows: 

\begin{equation} \label{equ:12}
Clusters = HDBSCAN({SLOU}_{1}, \cdots, {SLOU}_{\tau})
\end{equation}

\begin{equation}
Preserve : \{i \in {Cluster}_{major}| {u}_i \}
\end{equation}

where the $Clusters$ is the clustering result by using HDBSCAN algorithm, and the ${Cluster}_{major}$ is the set of nodes' id in the largest cluster.

\textbf{AGGREGATION:} In the last step, the server aggregates the preserved $\tau^{'}$ local model updates to update the global model:

\begin{equation}
{u}^{t+1} = \sum_{i=1}^{\tau^{'}} {u}^{t}_i
\end{equation}

\begin{equation}
{w}^{t+1} = {w}^{t} + {\eta}_{\text{g}} {u}^{t+1}
\end{equation}

where ${w}^{t+1}$ is the aggregated global model for ${(t+1)}^{th}$ iteration and ${\eta}_{\text{g}}$ is the global learning rate. 

\subsection{Complete XMAM algorithm} 
Algorithm \ref{alg:xmam} represents our complete XMAM method in a specific global iteration. For the server, it needs to take the following three steps: (i) After receiving a certain number of client local model updates, the server inputs a matrix into the models, whose parameters are local model updates, to get the SLOUs. (ii) The server clusters these SLOUs by the HDBSCAN algorithm and sorts these clusters according to the number of SLOUs to get the major cluster. (iii) The server aggregates the local model updates whose SLOUs are in the major cluster to update the global model.

\begin{algorithm}
\normalem
	\caption{Illustration of \textbf{XMAM} on $\tau$ clients $C_1$, $C_2$, ···, $C_{\tau}$ in $t^{th}$ global iteration. $\tau^{'}$ is the number of preserved local model updates whose SLOUs in ${Cluster}_{major}$.}
	\label{alg:xmam}
    \KwIn{received $\tau$ client local model updates; a random matrix $\mathcal{D}_{mat}$.}
    \KwOut{the global model ${w}^{t+1}$.} 
	\BlankLine

	/*Step \uppercase\expandafter{\romannumeral1}: Examining*/
	
	\For{$i$ = $C_1$, $C_2$, ···, $C_{\tau}$}
	{
	    ${SLOU}_{i} = Network({u}^{t}_{i}, D_{mat})$ 
	} 
	
    /*Step \uppercase\expandafter{\romannumeral2}: Clustering */
    
	${Clusters} = HDBSCAN({SLOU}_{1}, \cdots, {SLOU}_{\tau})$ 
	
	${Cluster}_{major} = Sort({Cluster}_1, {Cluster}_2, \cdots)[0]$ 
	\BlankLine
	/*Step \uppercase\expandafter{\romannumeral3}: Aggregation*/
	
	${w}^{t+1} = {w}^{t} + {\eta}_{\text{g}} \sum_{i=1}^{\tau^{'}} {u}^{t}_i (\tau^{'} \in Cluster_{major})$
	
	$ return \quad  {w}^{t+1} $
\end{algorithm}

\subsection{Mathmatical explanations}

In this section, we illustrate our method in mathematical terms. Due to the network that we use in our experiments is Convolutional Neural Networks (CNN), we conduct a convolution layer, a pooling layer, and a fully connected layer to get the output.

\begin{center}
$\begin{bmatrix}i_{1,1} & \dots & i_{1,n} \\ \vdots &\ddots &\vdots \\ i_{n,1} & \dots & i_{n,n} \end{bmatrix}
\xrightarrow[step:1]{core size:3}
\begin{bmatrix}A_{1,1} & \dots & A_{1,n-2} \\ \vdots &\ddots &\vdots \\ A_{n-2,1} & \dots & A_{n-2,n-2} \end{bmatrix}$\newline
$(core:\begin{bmatrix}a_{1,1} & a_{1,2} & a_{1,3} \\ a_{2,1} &a_{2,2} &a_{2,3} \\ a_{3,1} & a_{3,2} & a_{3,3} \end{bmatrix},A_{p,q}=i_{p-1,q-1}*a_{3,3}+i_{p,q-1}*a_{3,2}+\dots+i_{p+1,q+1}*a_{1,1})$\newline

$Relu=Relu\Bigg(\begin{bmatrix}A_{1,1} & \dots & A_{1,n-2} \\ \vdots &\ddots &\vdots \\ A_{n-2,1} & \dots & A_{n-2,n-2} \end{bmatrix}+b\Bigg)=Relu\Bigg(\begin{bmatrix}A_{1,1}+b & \dots & A_{1,n-2}+b \\ \vdots &\ddots &\vdots \\ A_{n-2,1}+b & \dots & A_{n-2,n-2}+b \end{bmatrix}\Bigg)$

$Pool=maxpooling(Relu)$ \ \ (kernel size=3)\newline

$output=\begin{bmatrix}P_{1,1} & \dots & P_{1,n-4} \\ \vdots &\ddots &\vdots \\ P_{n-4,1} & \dots & P_{n-4,n-4} \end{bmatrix}\times
\begin{bmatrix}s_{1}\\ \vdots \\ s_{n-4} \end{bmatrix}
+\begin{bmatrix}\hat{b_{1}}\\ \vdots \\ \hat{b_{n-4}} \end{bmatrix}
=
\begin{bmatrix}\sum{P_{1,i}s_{i}}+\hat{b_{1}}\\ \vdots \\ \sum{P_{n-4,i}s_{i}}+\hat{b_{n-4}} \end{bmatrix}
=
\begin{bmatrix}out_{1}\\ \vdots \\ out_{n-4} \end{bmatrix}$\newline

$SLOU=softmax\Bigg(\begin{bmatrix}out_{1}\\ \vdots \\ out_{n-4} \end{bmatrix}\Bigg)=\begin{bmatrix}slou_{1}\\ \vdots \\ slou_{n-4} \end{bmatrix}$\newline
\end{center}

This is the whole process of the random matrix passing through the network that we design for giving an example. We will explain why this process can distinguish malicious local model updates from benign ones in \ref{app:proof}. 

\section{Adaptive attacks} \label{section:Adaptive}
The adversaries may design adaptive attacks to bypass the detection after knowing the aggregation method used in the FL system. Adaptive attacks aim to increase the testing error rate of the global model. However, adaptive attacks require more knowledge about the current FL system. First of all, a malicious client needs to know which aggregation method is in use, which might not be public knowledge.

To further test XMAM's defensive capability, we evaluate it on Krum attack \cite{fang2020local}, which is an adaptive attack designed for attacking Krum and Multi-Krum. Furthermore, we develop an adaptive attack, XMAM attack, according to the general framework proposed by \cite{fang2020local}, and evaluate our method on it.  

\subsection{A general adaptive attack framework}

The adaptive attack framework proposed by \cite{fang2020local} is general to all aggregation methods. In adaptive attacks, the malicious clients collude to cause a deviation of the global model within the detective boundary. The most effective deviation is to find the opposite direction of the global model update and then modify the local model update to this direction. Except for the direction, the remaining metric to think about is the magnitude. Therefore, how to find the maximum magnitude within the detective boundary is critical. The general adaptive attack framework can be defined as follows:

\begin{equation}
\begin{split}
    &\max_{\lambda \in R^+} \lambda \\
    Subject \quad to \quad {u_{1}^{'}} &= \mathcal{A} ( {u}_{1}^{'}, \ldots, {u}_{f}^{'}, {u}_{f+1}, \ldots, {u}_{\tau} ),\\
    {u}_{1}^{'} &= {u}_{\textit{g}} - \lambda {s},\\
    {u}_{i}^{'} &= {u}_{1}^{'},  (i = 2,3,\ldots,f).
\end{split}
\end{equation}

where ${u}_{1}^{'}, \ldots, {u}_{f}^{'}$ are the $f$ malicious local model updates, $\mathcal{A}()$ is the aggregation method that the malicious clients want to attack, ${s}$ is the $Sign()$ of the global model update and $\lambda$ is the magnitude that maximizes the bounded attack effect. 

\vspace{-2mm}
\subsection{Threat model for our adaptive attack}
We assume all malicious clients are collusive, and their leader can obtain all client local model updates and arbitrarily modify the malicious local model updates. After receiving the global model, these malicious clients train their local model updates with their clean local data. We use these local model updates and other benign local model updates to form a distribution. Then, the leader searches for a suitable $\lambda$ according to the distribution to bypass detection and modifies the other malicious client local model updates uniformly. This threat model is in line with the full knowledge assumption \cite{fang2020local,cao2020fltrust}. 
\vspace{-2mm}
\subsection{Our complete adaptive attack algorithm}
We set initial $\lambda=1$. As in \cite{fang2020local}, we use a binary search to find the ultimate $\lambda$. Specifically, we first calculate the correct direction of global update $s$, which is the $Sign(\sum_1^{\tau} u_i)$. The function $Sign()$ has been illustrated in Section \ref{oth:rsa}. Then, we set the malicious local model updates as the global update in the previous iteration and deviate it to the inverse direction $s$ in a certain magnitude $\lambda$. 
If XMAM cannot catch the deviated malicious local model updates, we will reduce the magnitude $\lambda$ by half. Otherwise, we will return the current magnitude $\lambda$. That means the returned magnitude $\lambda$ is the maximum attack magnitude that the attackers can implement.

\begin{algorithm}
\normalem
	\caption{XMAM attack}
	\label{alg:xmam-attack}
    \KwIn{the global update ${u}_{\textit{g}}$; the $\tau$ client local model updates; a random matrix $\mathcal{D}_{mat}$.}
    \KwOut{the deviation parameter $\lambda$.} 
	\BlankLine
    Initialize $\lambda$ = 1
    
    \While{$C_1$, $C_2$, ···, $C_{f}$ $\notin{{Cluster}_{major}}$ $or$ $\lambda$ $\le$ $1e^{-10}$}
    {
        % $ $\lamuda$ = \frac{$\lamuda$}{2}$
        
        \For{$i$ = $C_1$, $C_2$, ···, $C_{f}$}
    	{
    	    ${u}_{i}$ = ${u}_{\textit{g}}$ - $ \lambda {s}$ 
    	}
    	
    	${Clusters} = HDBSCAN({SLOU}_{1}, \cdots, {SLOU}_{\tau})$ 
    	
    	${Cluster}_{major} = Sort({Cluster}_1, {Cluster}_1, \cdots)[0]$ 
    }
	$return$ \quad $\lambda$ 
\end{algorithm}

\section{Evaluation}

In this section, we first evaluate the fidelity of XMAM and the other six aggregation methods. Then, we implement the three backdoor attacks to FL with and without hiding techniques. we compare the six former aggregation methods with XMAM to highlight the superior capability of XMAM for defending against backdoor attacks. Furthermore, we implement the adaptive attack (Krum attack) proposed by \cite{fang2020local} and self-designed adaptive attack (XMAM attack) to demonstrate the robustness of XMAM. At last, we theoretically analyze the time complexity of each aggregation method and test the real run time of one-round aggregation on our device. The results show that our method is efficient. For example, our method is about 10000 times faster than Multi-Krum. For simplicity, in our experiments, we use an all-ones matrix to evaluate our method. Our code is publicly available at \textit{https://github.com/BESTICSP/XMAM}.

\subsection{Experimental setup}

1) \textit{Datasets}: We use two datasets in the computer-vision domain and sample them to simulate non-i.i.d. data. Note that we do not conduct our experiments in the i.i.d. scenario since the malicious local model updates are much easier to be detected in this scenario. As in \cite{wang2020attack}, for each dataset, we divide the training data into $N$ piles using Dirichlet distribution\cite{hsu2019measuring} and allocate them to $N$ clients to simulate the practical FL system (\textit{i.e.}, the data distribution is non-i.i.d., and $\chi \sim Dir(0.5, N)$, where $\chi$ is the data distribution, $N$ is the number of total clients, and $0.5$ is the distribution parameter.) 

\textbf{CIFAR-10 \cite{krizhevsky2009learning}:} CIFAR-10 is a color image classification dataset incorporating predefined 50,000 training examples and 10,000 testing examples. Each example belongs to one of the 10 classes (airplane, automobile, bird, cat, deer, dog, frog, horse, ship, and truck).

\textbf{MNIST \cite{lecun1998gradient}:} The MNIST dataset of handwritten digits has a training set of 60,000 examples and a test set of 10,000 examples. It is a subset of a larger set available from NIST. The digits have been size-normalized and centered in a fixed-size image.

2) \textit{Evaluated backdoor attacks and adaptive attacks}: We first evaluate the performance of the seven aggregation methods (FedAvg, Krum, Multi-Krum, NDC, RFA, RSA, and XMAM) on the three backdoor attacks (Trigger attack, Semantic attack, and Edge-case attack). To show the threat of the backdoor attacks with hiding techniques and the robustness of XMAM, we respectively conduct experiments on Black-box mode, PGD mode, and SMP mode. Then, we evaluate our method for adaptive attacks. Note that the three attack modes are designed for backdoor attacks to ensure that the server will select the malicious local model updates. We do not conduct adaptive attacks on them since the adaptive attacks have achieved this.

\textbf{Trigger attack:} As in \cite{gu2017badnets}, we poison 30\% local data of malicious clients by stamping a square pixel block on the corner of images. 

\textbf{Semantic attack:} As in \cite{bagdasaryan2020backdoor}, we use the images of green cars (class: 1) to generate the poisoned training data and backdoor test dataset. Consistent with \cite{bagdasaryan2020backdoor}, we set the target label as the bird (class: 2).

\textbf{Edge-case attack:} As in \cite{wang2020attack}, we use the images of the southwest airplane, which is in the tail of data distribution, to generate the poisoned training data and backdoor test dataset, and set the target label as the truck (class: 9). Note that the images of the southwest airplane would be classified as an automobile (class: 1) if we do not implement backdoor attacks.

\textbf{Adaptive attack:} We use the adaptive attack designed in Section \ref{section:Adaptive}. We assume the server collects 20 clients in each iteration, where 5\% to 50\% of them are malicious clients.

\textbf{Attack modes:} For Black-box mode, a malicious client trains his local model in the poisoned dataset and submits the local model update to the server. For PGD mode, we set the PDG parameter $\epsilon=5e^{-2}$ for three backdoor attacks. For SMP mode, as in \cite{bhagoji2019analyzing}, we set ${\rho}_1=10$ and ${\rho}_2=1e^{-4}$.

% since if we set it more little, the effect of backdoor attacks might be lose (Fig. \ref{fig:eps})
% \begin{figure}[ht]
% \begin{center}
% \subfigcapskip = -5pt
% \subfigure[Sign]{
%     \includegraphics[width=0.6\linewidth]{eps.pdf}
% }
% \vspace{-3mm}
% \caption{The test accuracy of XMAM and Multi-Krum under Sign-flipping attack and Same-value attack on CIFAR-10 and EMNIST. There are 9 poisoned models out of 20 collected ones in each iteration (45\% poisoned models).} \vspace{-6mm}
% \label{fig:eps}
% \end{center}
% \end{figure}

3) \textit{Defenders' setting}: As in \cite{wang2020attack}, in our experiments, we set $\delta=2$ for NDC, and we set the smoothing factor $v=0.1$, the fault tolerance threshold $\mu=10^{-5}$, and the maximum number of rounds $R=500$ for RFA. We set the learning rate of RSA $\beta_r = 5 \times 10^{-5} \times 0.998^t $, which is proved by experiments that in this setting, and RSA performs well when there is no attack. 

4) \textit{Evaluation metrics}: Consistent with \cite{cao2020fltrust}, we use attack success rate to reflect the performance of backdoor attacks since the aim of them is to promote the accuracy of the backdoor task. Specifically, the attack success rate is the fraction of data in the backdoor task, which is classified by the global model as the attacker-chosen class. For adaptive attacks, we use testing error rates to reflect the performance of adaptive attacks since their goal is to increase the testing error rate of the testing dataset. Specifically, the testing error rate is the fraction of data in the testing dataset that are mistakenly classified by the global model.

5) \textit{System settings}: In the experiments of backdoor attacks, we set 200 clients in the simulated FL system, the server collects 30 clients in each iteration, and 20\% of the 30 clients are malicious clients. In the experiments of adaptive attacks, we set 200 clients in the simulated FL system, the server collects 30 clients in each iteration, and 40\% of the 30 clients are malicious clients. Furthermore, we plot the PCA scatter diagram in the setting that the server collects 100 clients in each iteration, and 20\% of the 100 clients are malicious clients. For CIFAR-10, we use network VGG9, and for MNIST, we use network LeNet. See Table \ref{tab:setting} for more details.

\begin{table*}
\setlength{\tabcolsep}{0.5mm}{
\begin{center}
\caption{The default FL system parameter settings.} \vspace{-3mm}
\label{tab:setting}
\centering
\begin{tabular}{|c|c|c|c|c|c|c|c|}
\hline
                                    & \multicolumn{3}{c|}{Backdoor attacks} & \multicolumn{2}{c|}{Adaptive attacks}        \\ \hline
                                    & Trigger attack   & Semantic attack  & Edge-case attack    & Krum attack  & XMAM attack                    \\ \hline
 Total number of clients            & \multicolumn{5}{c|}{200}                                                               \\ \hline
 Clients selected in each iteration & \multicolumn{5}{c|}{30}                                                                 \\ \hline
 Byzantine clients in each iteration   & \multicolumn{3}{c|}{20\%}                           & \multicolumn{2}{c|}{40\%}                         \\ \hline
 Frequency of attacks               & \multicolumn{5}{c|}{1}                                                                            \\ \hline
 Local iterations                   & \multicolumn{5}{c|}{1}                                           \\ \hline
 Global iterations                  & \multicolumn{3}{c|}{100}                           & \multicolumn{2}{c|}{50}                       \\ \hline
 Batch size                         & \multicolumn{5}{c|}{32}                                                                           \\ \hline
 Combined learning rate             & \multicolumn{5}{c|}{$0.001 \times 0.998^t$}                                                       \\ \hline
 Optimizer                           & \multicolumn{5}{c|}{SGD}                                                                            \\ \hline
 Momentum                           & \multicolumn{5}{c|}{0.9}                                                                            \\ \hline
 Weight decay                       & \multicolumn{5}{c|}{$10^{-4}$}                                                                           \\ \hline
   
\end{tabular}
\end{center}
}
\end{table*}

\subsection{Experimental results} \label{sec:exp}

\begin{figure}[ht]
\begin{center}
\includegraphics[width=0.35\linewidth]{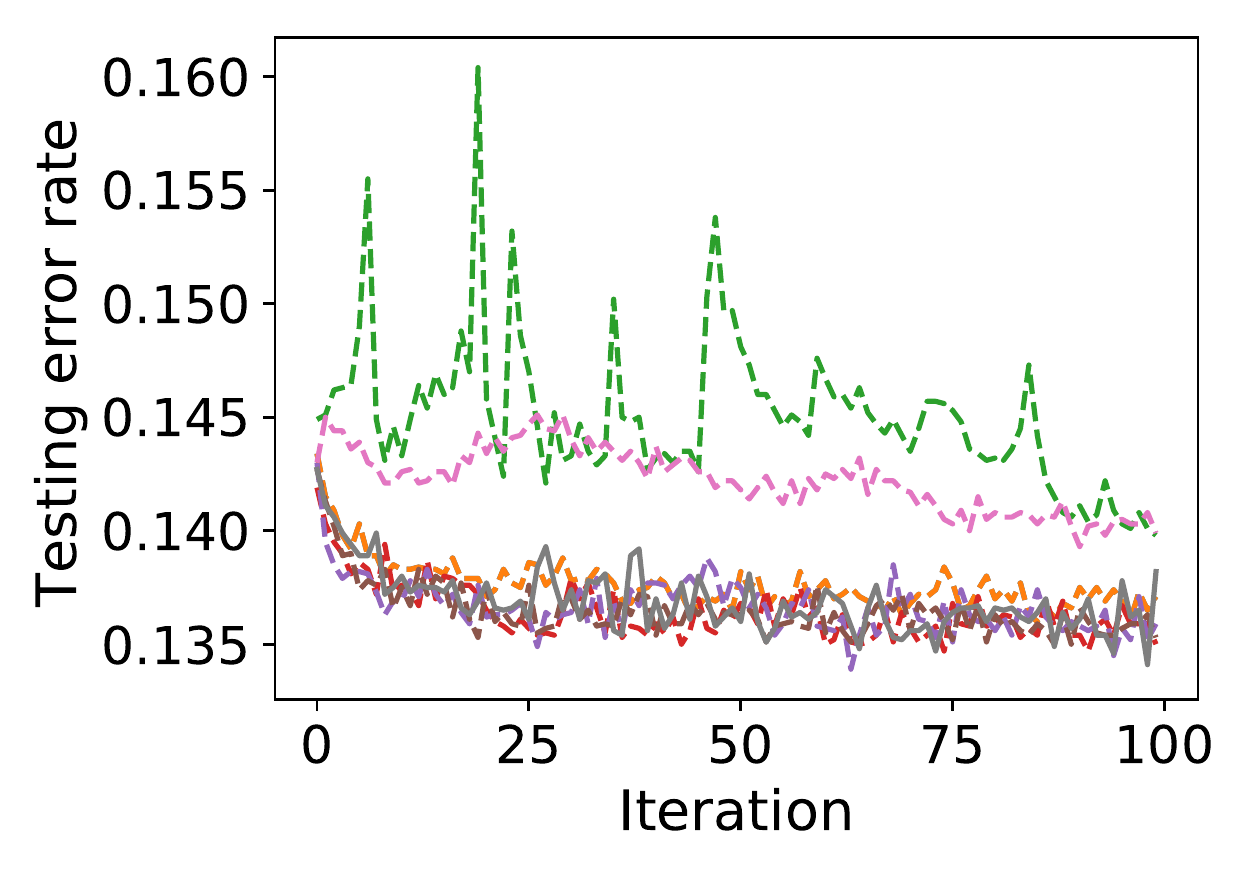}
\end{center}
\begin{center}
\includegraphics[width=0.3\linewidth]{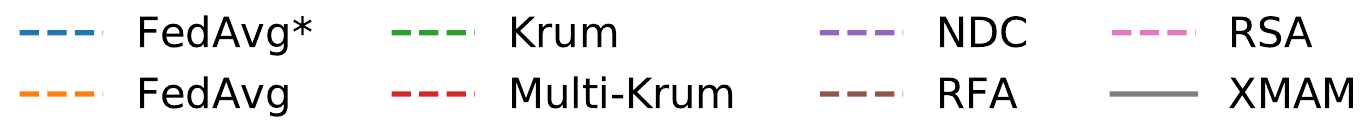}
\end{center}\vspace{-3mm}
\caption{The testing error rate of the global model under different aggregation methods when there is no attack. We can find that Krum is fluctuant and has a high testing error rate, RSA has a gap to the FedAvg, and other methods have comparable performance to FedAvg.} \vspace{-4mm}
\label{fig:noattack}
\end{figure}

1) \textit{Fidelity}: When there is no malicious client in FL, we can see from Fig. \ref{fig:noattack} that all aggregation methods except Krum and RSA have a similar performance to FedAvg. That is to say, most existing methods can ensure fidelity. It is no surprise that Krum has such a fluctuation in testing error rate since it only collects one local model update used for the global model's updating in each iteration. To punish the malicious local model updates, RSA limits the magnitude of all received local model updates to a fixed number and only preserves the direction, which makes the ultimate global model tend to a sub-optimal solution.
%---------------------------------------- no attack

2) \textit{Robustness}: Viewing Fig. \ref{fig:backdoor_exp} horizontally, we can find that different backdoor attacks have different performances. In contrast, Trigger backdoor attack is harder to succeed. In black-box mode, Trigger backdoor attack only evades FedAvg and RFA, but Semantic and Edge-case backdoors both evade FedAvg, NDC, RFA, and RSA. This phenomenon is more obvious in the SMP mode. Why does Trigger backdoor attack have a poor capability on attack success rate when the poisoned data proportion is the same as the other two backdoor attacks? We conjecture the reason is that it is harder for a model to learn the feature of  the backdoor pattern from variant backgrounds than from a fixed background. Since Trigger backdoor attack poisons data by printing backdoor patterns on random data, but the backdoor patterns of Semantic and Edge-case backdoor attack are the data (containing special feature) itself, the success of Trigger backdoor attack needs more intensity and iterations. 

Viewing Fig. \ref{fig:backdoor_exp} vertically, we can observe the different performances on different attack modes. Visually, we can see that the attacks using hiding techniques successfully evade more aggregation methods than the attacks without using hiding techniques. Most notably, Multi-Krum is robust in black-box mode but fails in PGD and SMP mode (see Fig. \ref{fig:backdoor_exp}'s (b)(c)(e)(f)(h)). That means when backdoor attacks are under PGD mode and SMP mode, except Krum and our method, no existing aggregation methods can defend against them. About Krum, we will analyze it as follows. To illustrate why Multi-Krum fails but XMAM succeeds when the malicious local model updates are more concealing, we plot the distribution of 100 client local model updates using PCA dimension reduction technology. From Fig. \ref{fig:pgd_eps}, we can observe that with the scaling down magnitude increasing, the distance between malicious local model updates and benign local model updates are increasingly close, which makes them indistinguishable. In contrast, the SLOUs of malicious local model updates and benign local model updates still have a clear boundary when the malicious local model updates and benign ones are indistinguishable (Fig. \ref{fig:pgd_eps}(c)(f)). Since it collects $n-f$ local model updates that have a closer Euclidean distance to other local model updates, Multi-Krum fails when the scaling down magnitude is large enough.

Fig. \ref{fig:backdoor_main} shows the testing error rate of the global model on the main task. We can easily observe that whatever modes the backdoor attacks are under, our aggregation method ensures a low testing error rate, which is comparable to FedAvg*. However, other methods, especially RSA, have a higher testing error rate compared with XMAM.

\begin{figure}[ht]
\begin{center}  

\includegraphics[width=0.3\linewidth]{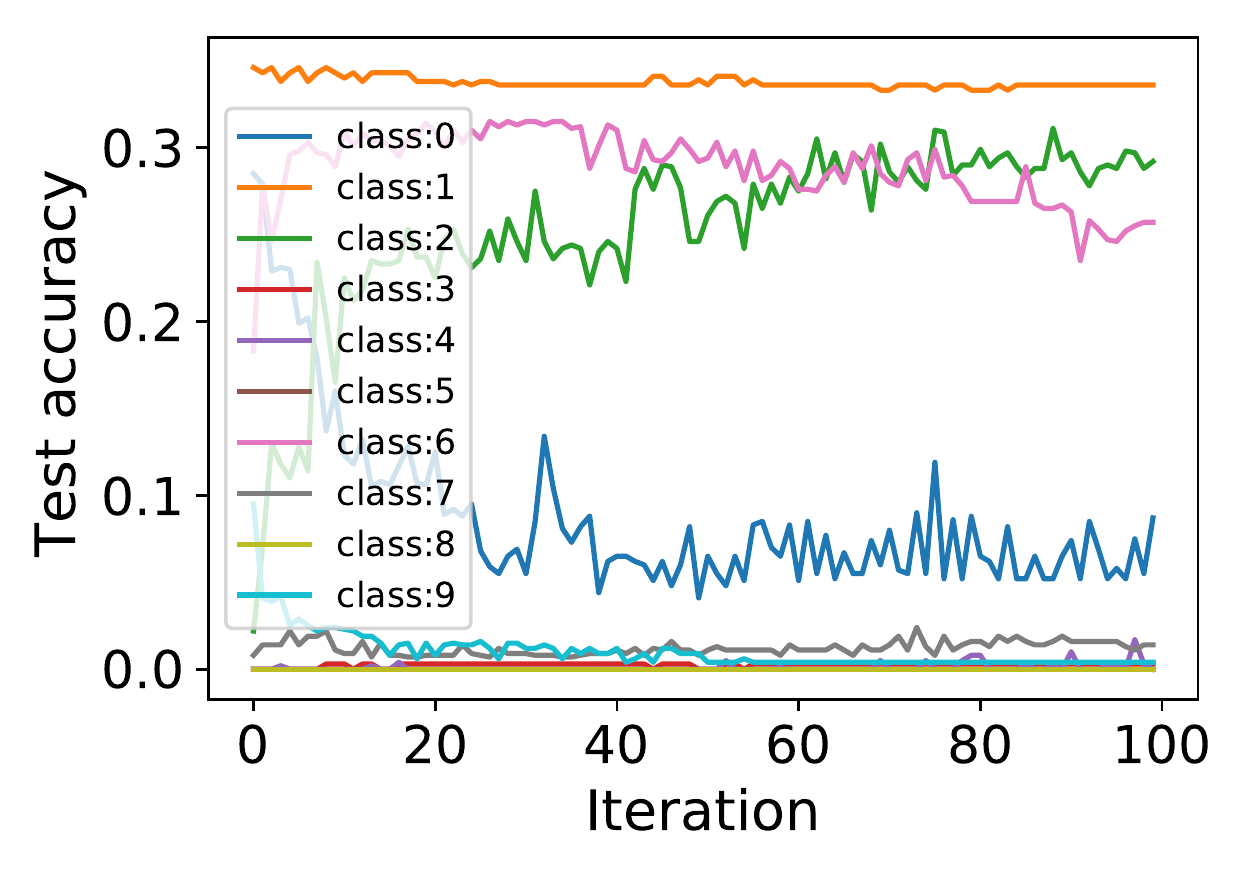}

\caption{The test accuracy of the global model on the backdoor testing dataset. We guess that the data of green cars are close to data of birds in the output space of the clean global model. So, we label the testing data as classes 0-9 to observe the test accuracy. The results show that the data of green cars are more likely to be classified as automobile (class 1: 33.6\%), bird (class 2: 29\%), and frog (class 6: 25.2\%), which verifies our conjecture.} \vspace{-4mm}
\label{fig:analysis}
\end{center}
\end{figure}

\begin{figure*}[ht]
\begin{center}
\subfigcapskip = -8pt
\subfigure[updates \& no scaling]{
  \includegraphics[width=0.26\linewidth]{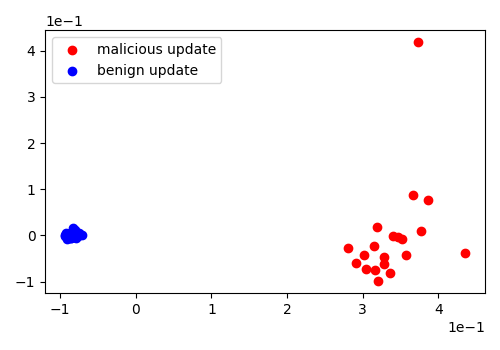}
}
\subfigure[updates \& $\epsilon$=5e-1]{
  \includegraphics[width=0.26\linewidth]{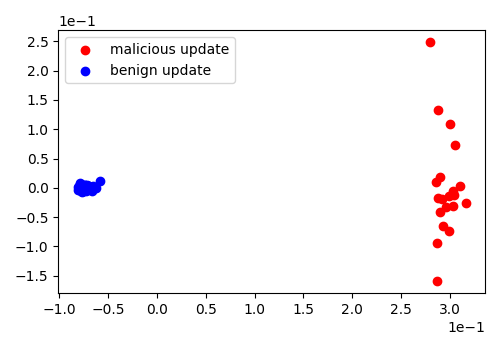}
}
\subfigure[updates \& $\epsilon$=5e-2]{
\includegraphics[width=0.26\linewidth]{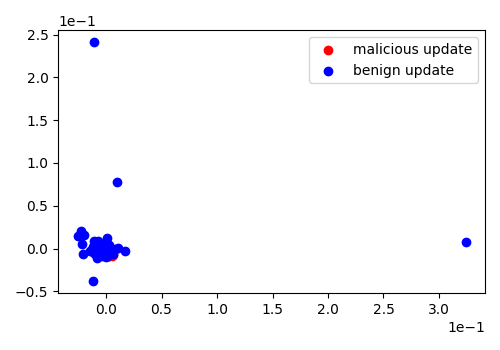}
}\vspace{-3mm}
\end{center}

\begin{center}
\subfigcapskip = -8pt
\subfigure[SLOUs \& no scaling]{
  \includegraphics[width=0.26\linewidth]{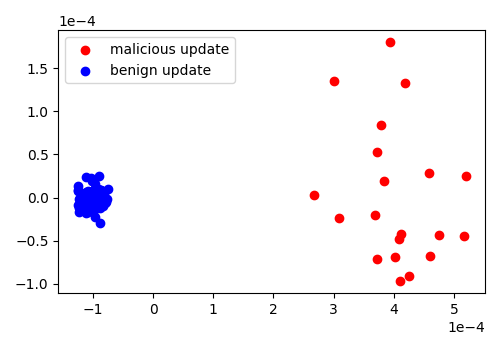}
}
\subfigure[SLOUs \& $\epsilon$=5e-1]{
  \includegraphics[width=0.26\linewidth]{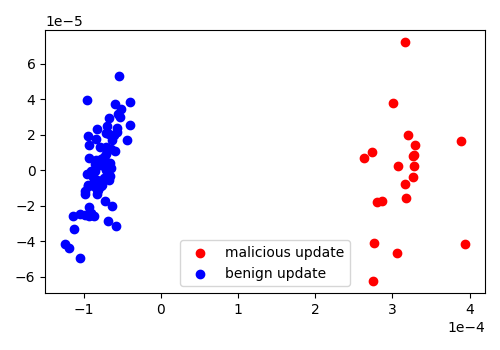}
}
\subfigure[SLOUs \& $\epsilon$=5e-2]{
  \includegraphics[width=0.26\linewidth]{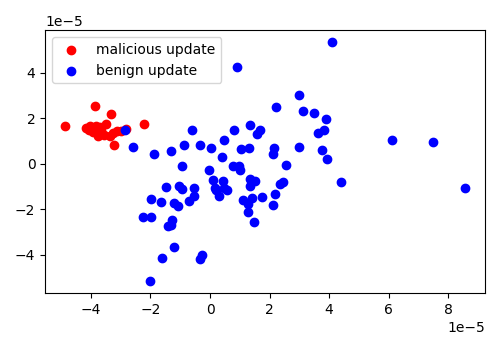}
}\vspace{-3mm}
\end{center}

\caption{The dimensionality reduction graph of the 100 local model updates and the 100 corresponding SLOUs. The red dots (20\%) are the malicious local model updates or SLOUs, and the blue dots (80\%) are the benign local model updates or SLOUs. From left to right, we successively decrease the PGD parameter $\epsilon$ (\textit{i.e.}, we scale down the local model updates to a smaller norm) to observe the distribution of the 100 local model updates and the 100 corresponding SLOUs. Thus, we can find that the PGD hiding technique makes the malicious local model updates indistinguishable from the benign local model updates, but can not hide the corresponding malicious SLOUs from benign SLOUs.} \vspace{-4mm}
\label{fig:pgd_eps}
\end{figure*}

\begin{figure*}[ht]
\begin{center}
\subfigcapskip = -8pt
\subfigure[Trigger \& Black-box]{
  \includegraphics[width=0.26\linewidth]{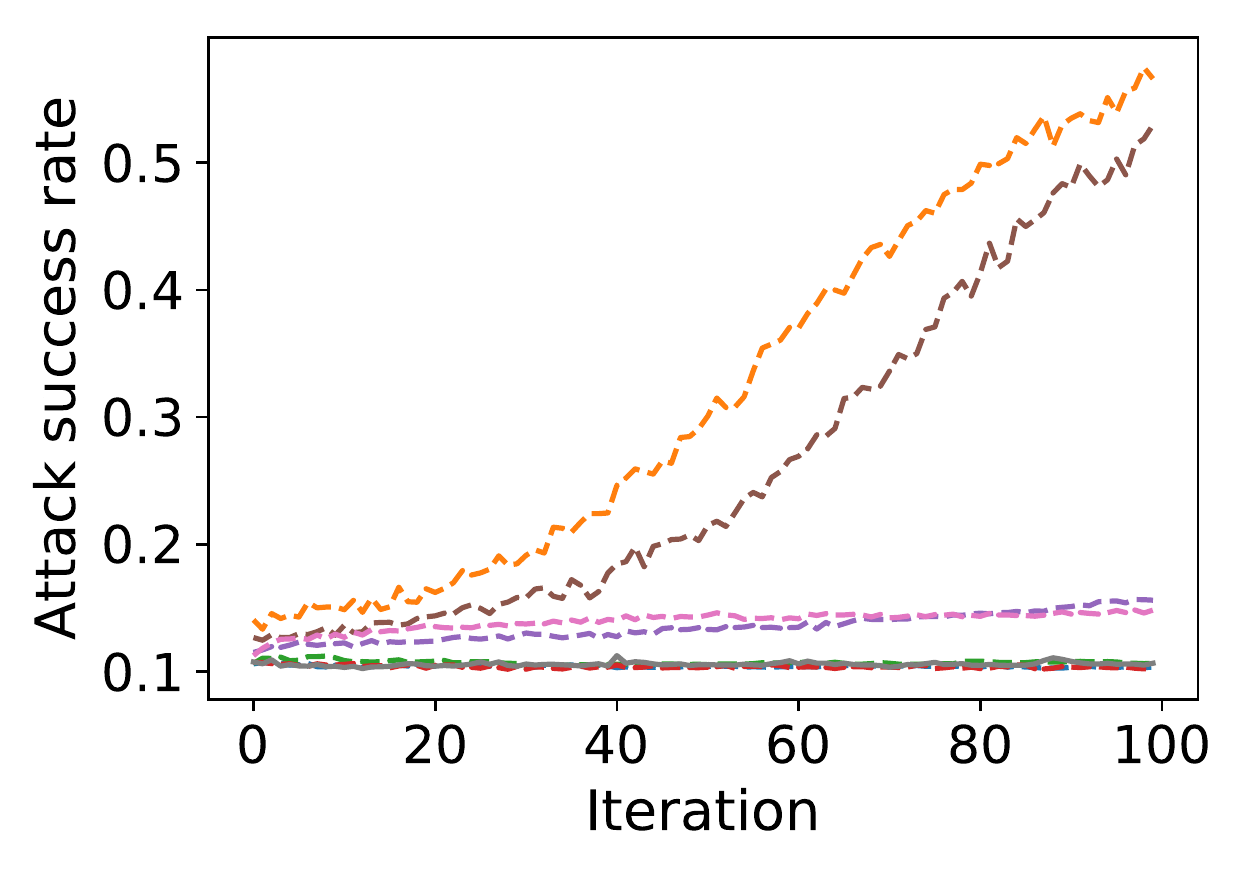}
}
\subfigure[Semantic \& Black-box]{
  \includegraphics[width=0.26\linewidth]{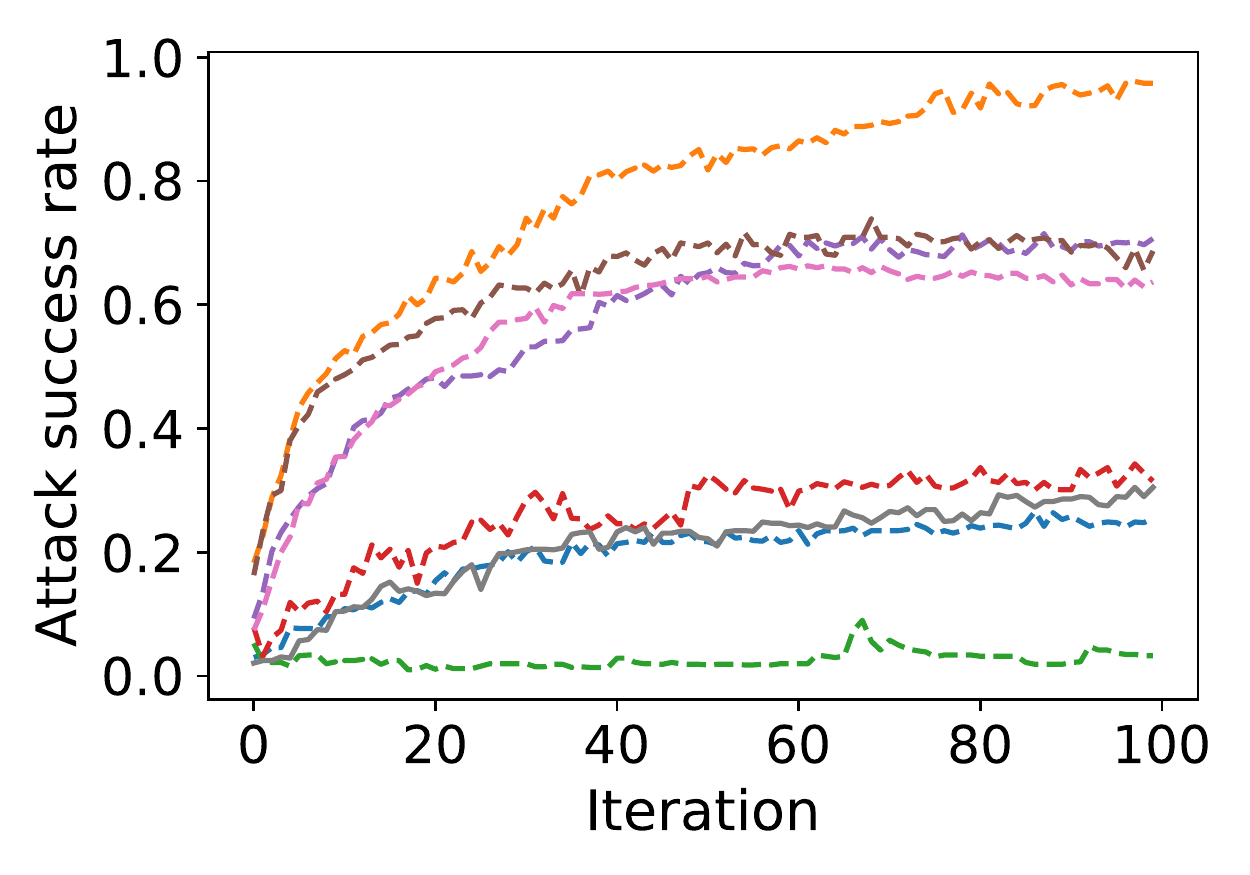}
}
\subfigure[Edge-case \& Black-box]{
\includegraphics[width=0.26\linewidth]{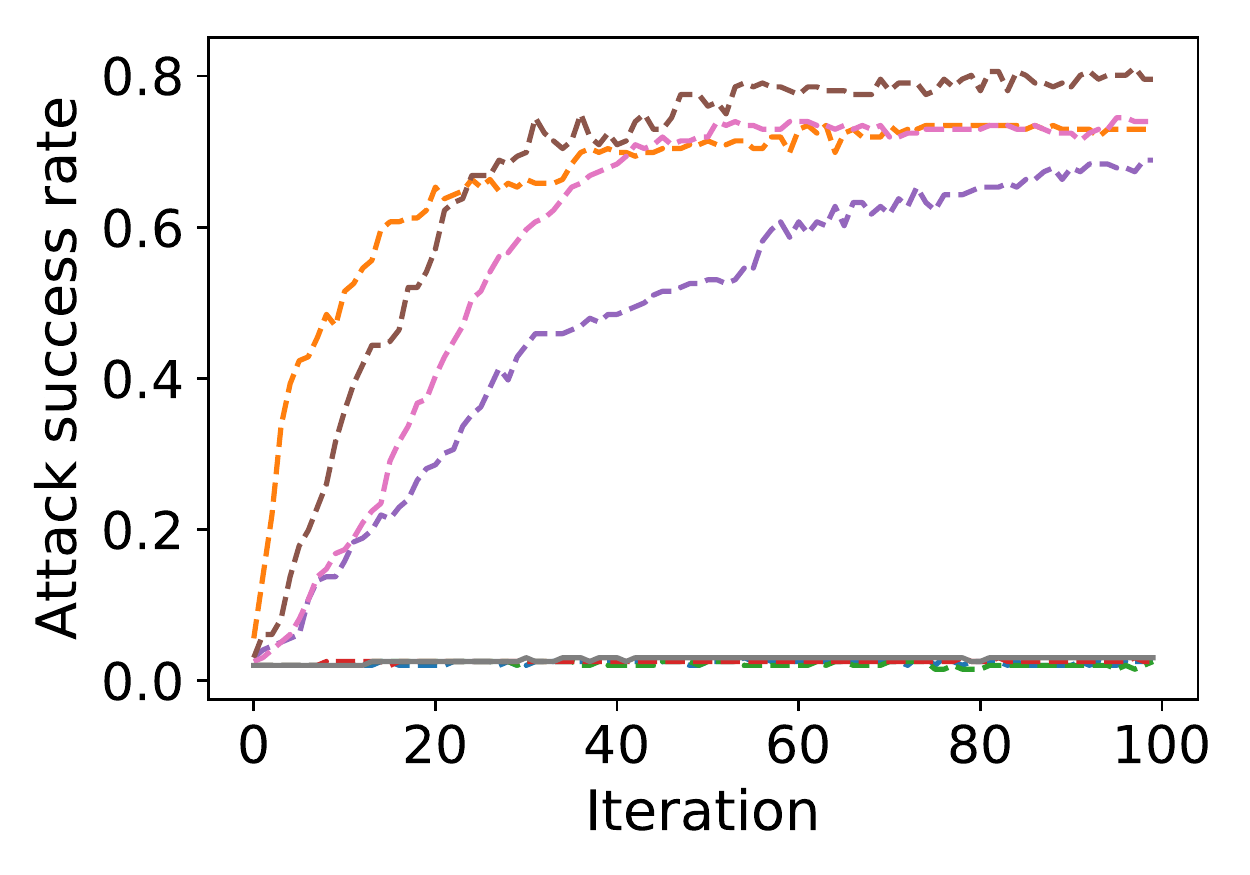}
}\vspace{-3mm}
\end{center}

\begin{center}
\subfigcapskip = -8pt
\subfigure[Trigger \& PGD]{
  \includegraphics[width=0.26\linewidth]{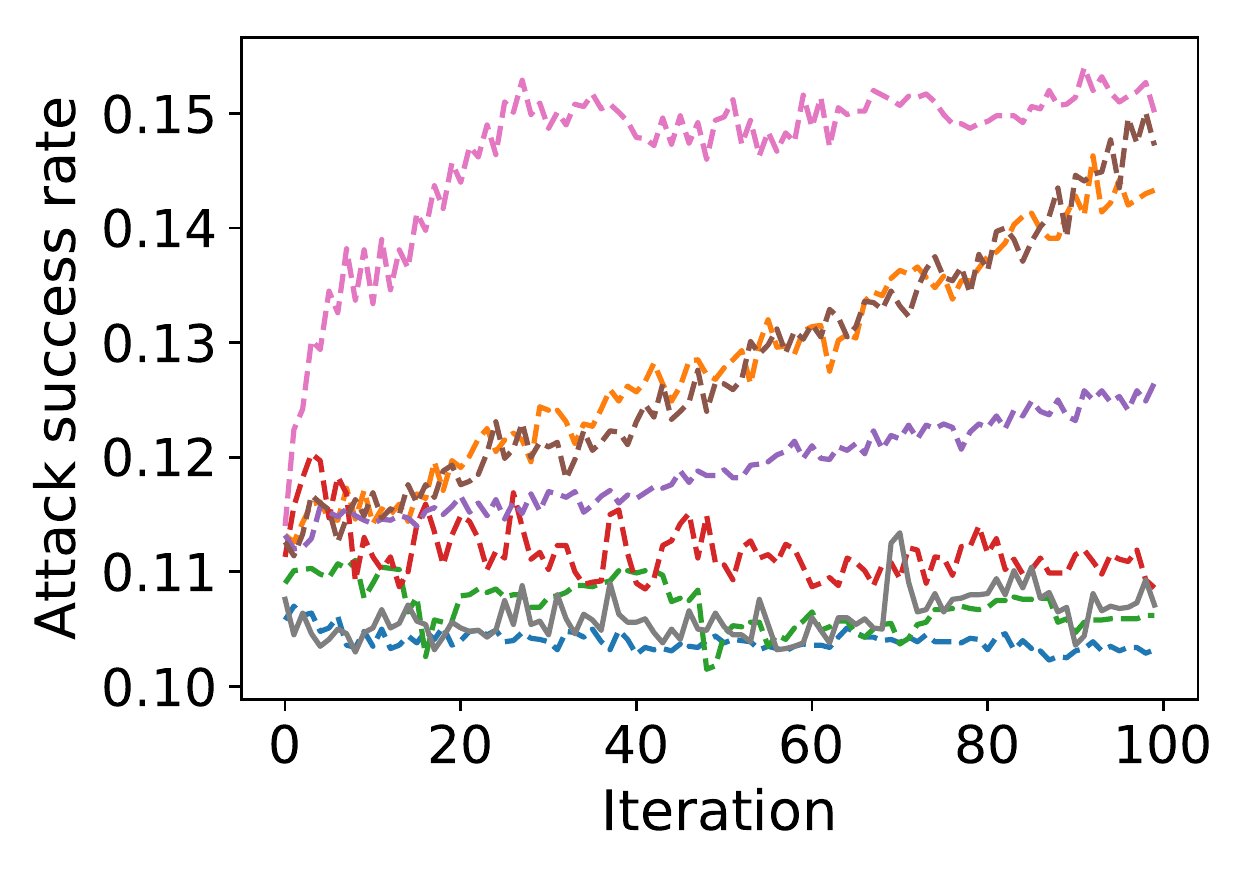}
}
\subfigure[Semantic \& PGD]{
  \includegraphics[width=0.26\linewidth]{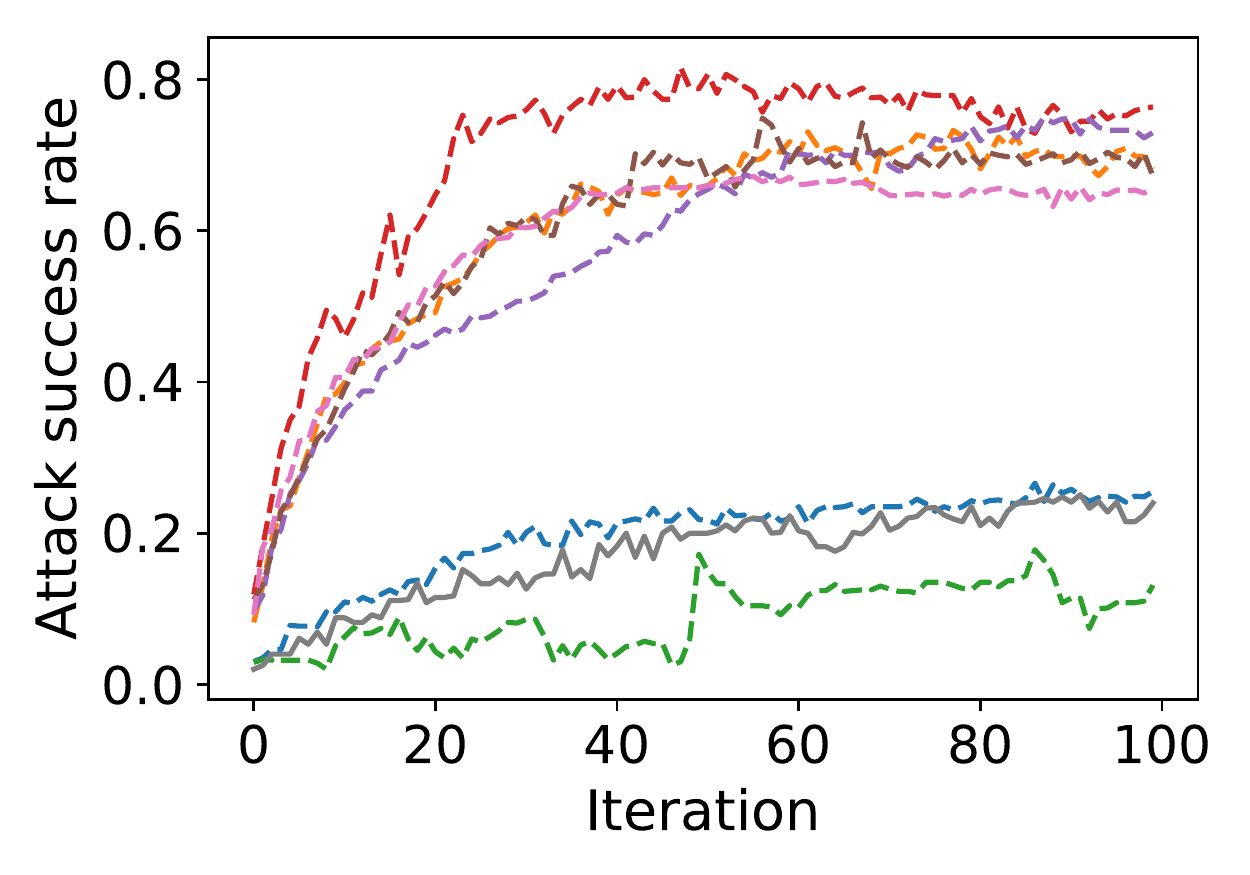}
}
\subfigure[Edge-case \& PGD]{
   \includegraphics[width=0.26\linewidth]{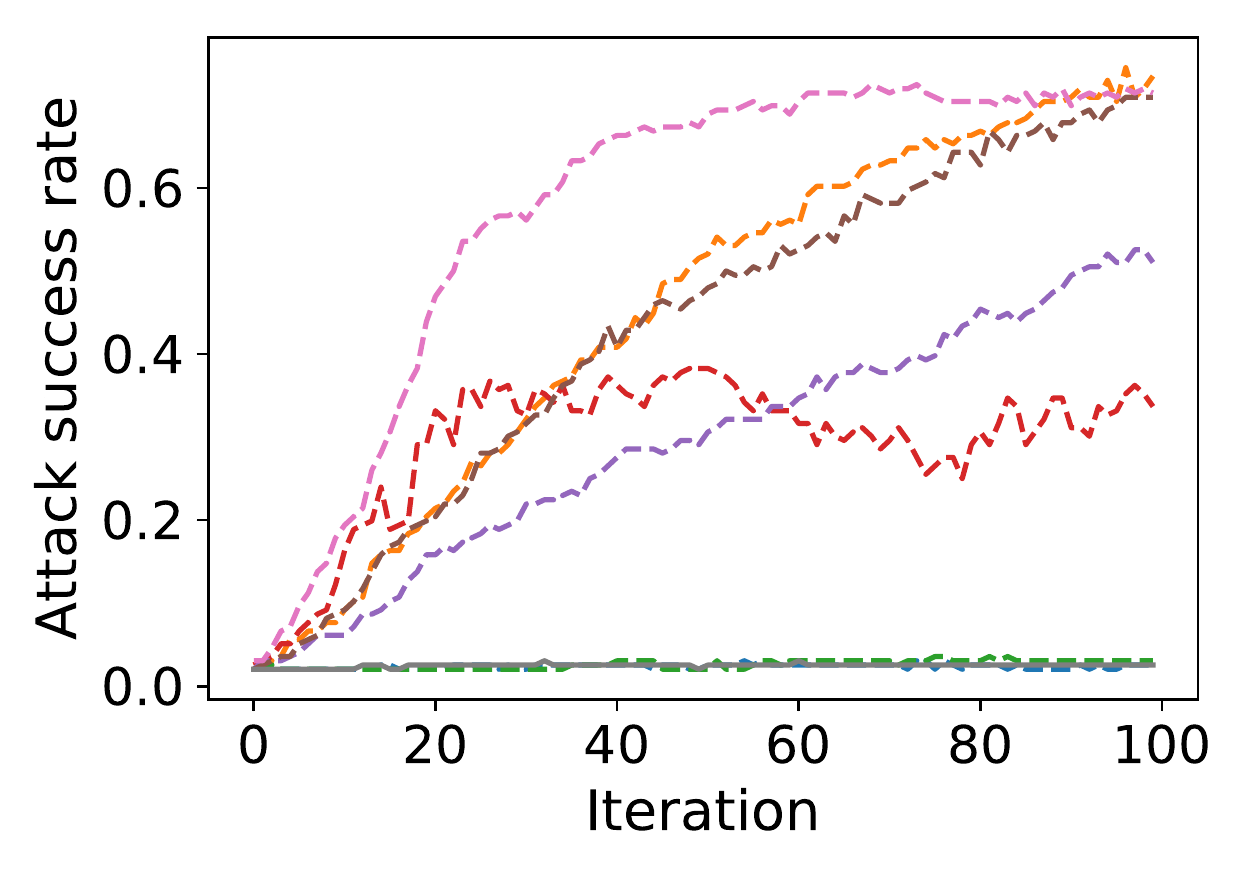}
}\vspace{-3mm}
\end{center}

\begin{center}
\subfigcapskip = -8pt
\subfigure[Trigger \& SMP]{
  \includegraphics[width=0.26\linewidth]{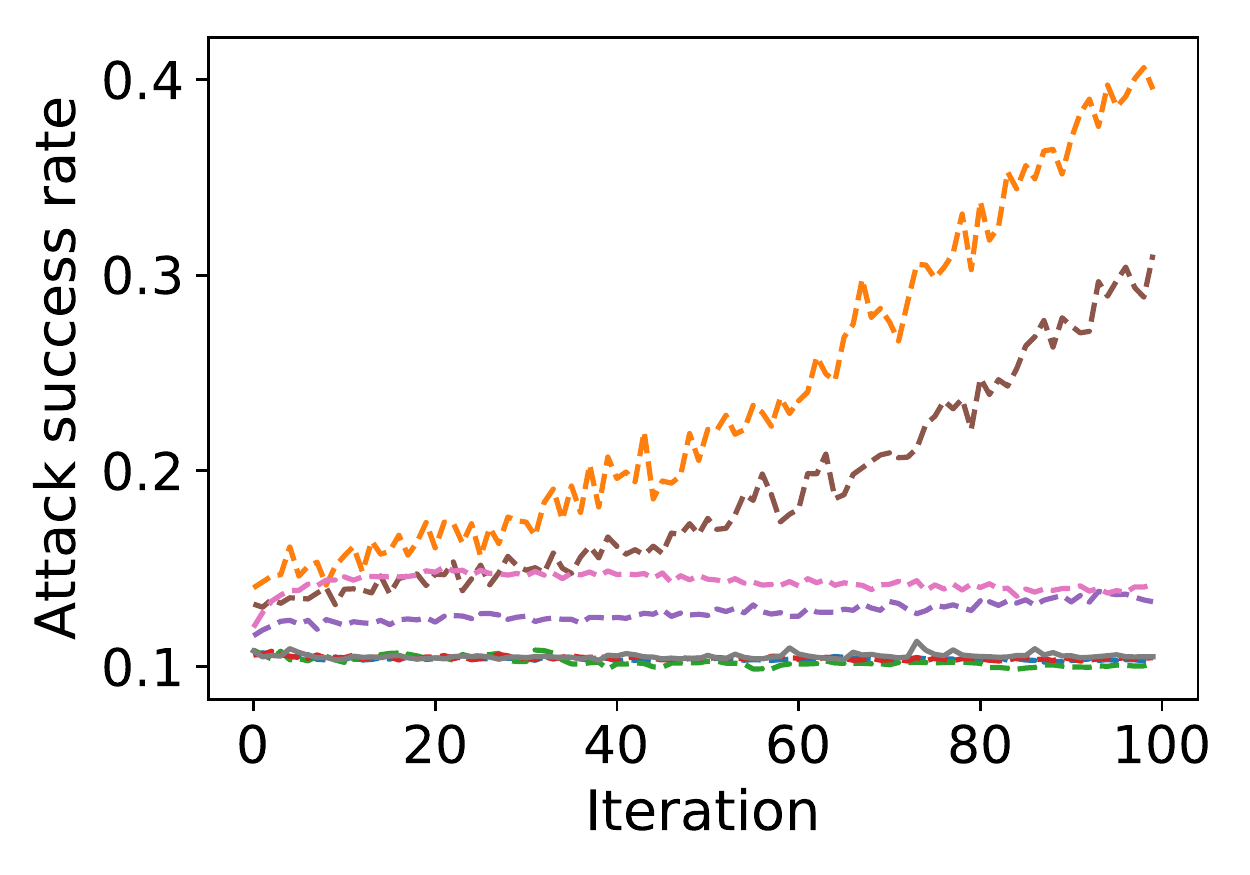}
}
\subfigure[Semantic \& SMP]{
  \includegraphics[width=0.26\linewidth]{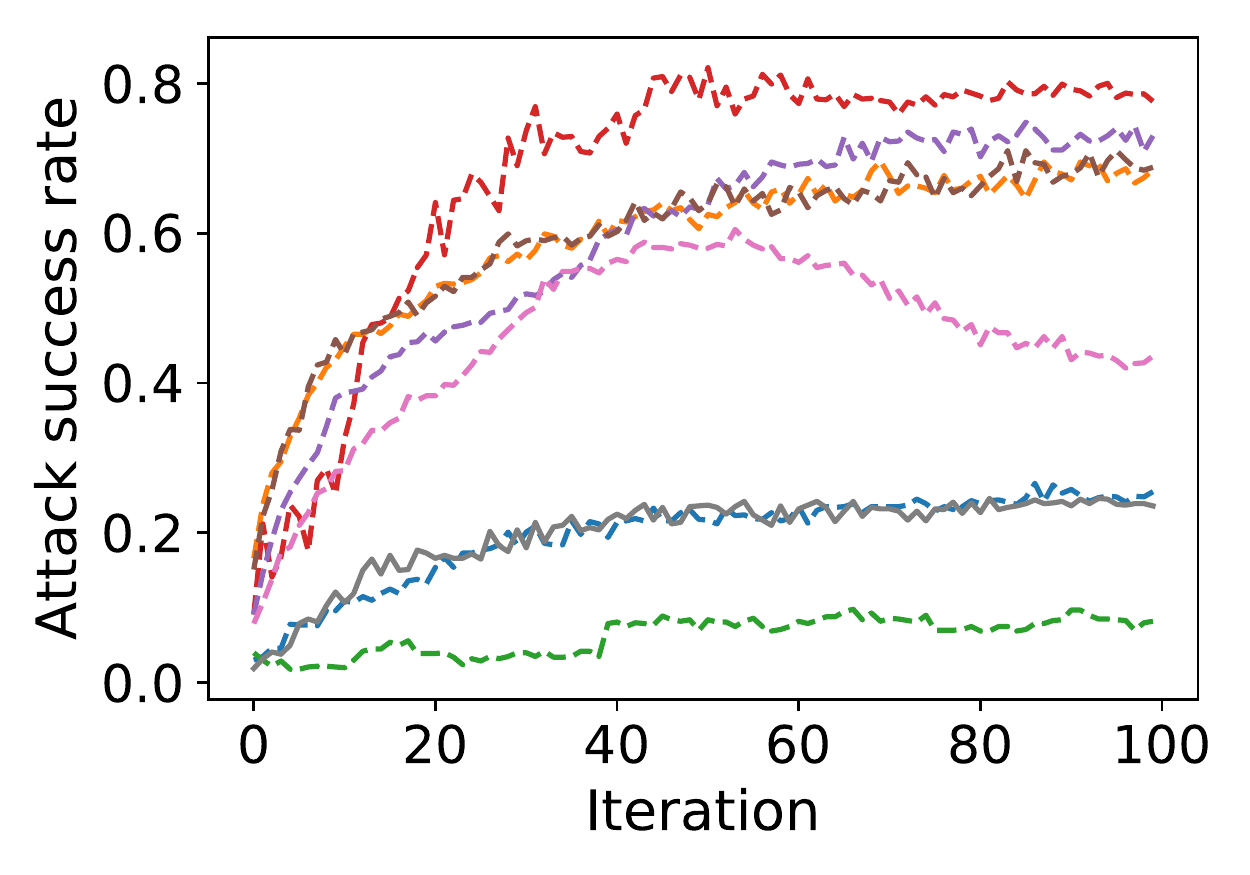}
}
\subfigure[Edge-case \& SMP]{
   \includegraphics[width=0.26\linewidth]{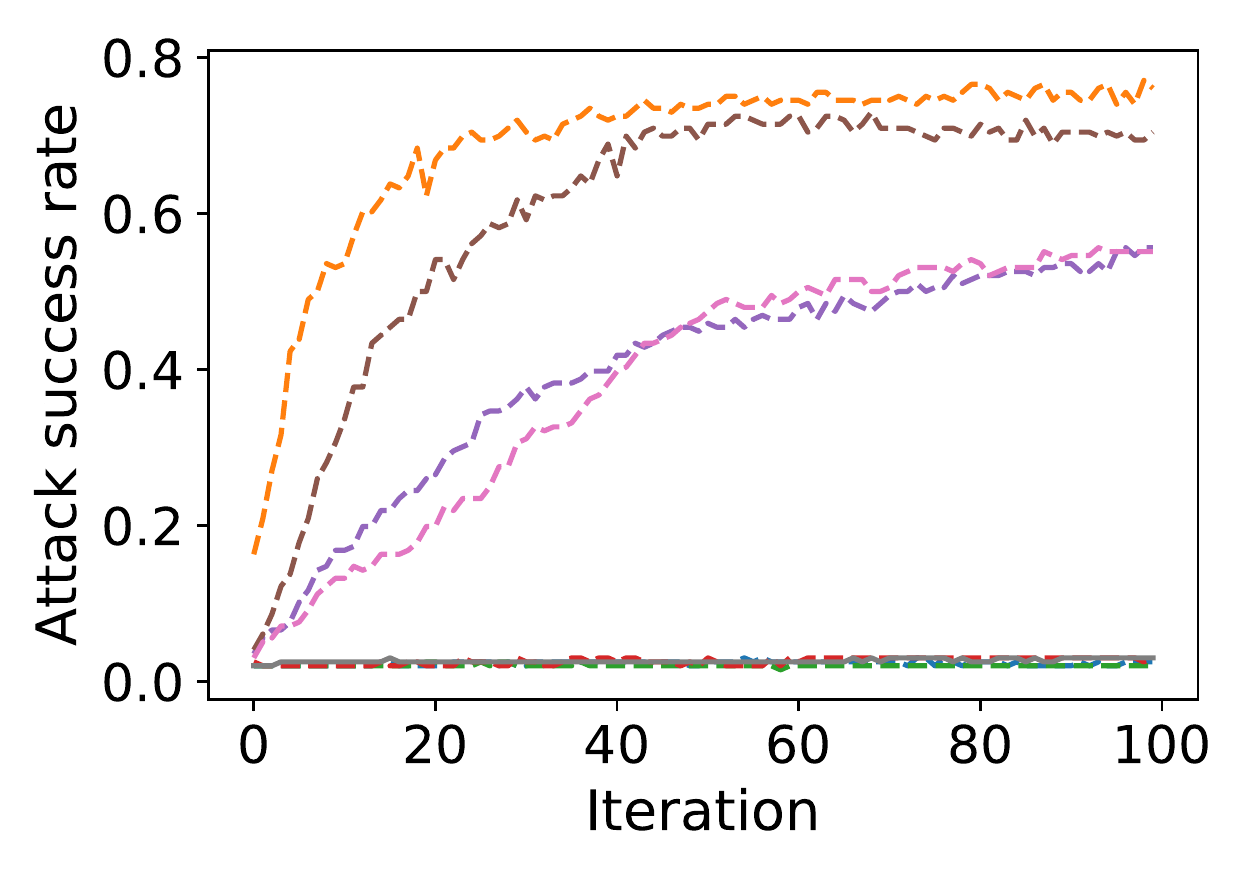}
}\vspace{-2mm}
\end{center}
\begin{center}
\includegraphics[width=0.8\linewidth]{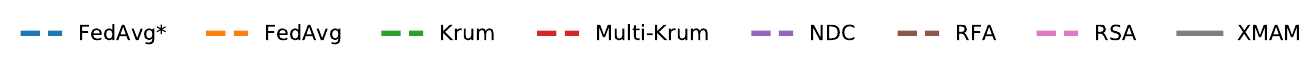}
\end{center}\vspace{-3mm}
\caption{The attack success rate of three backdoor attacks (Trigger attack, Semantic attack, and Edge-case attack) on the backdoor task under three modes (Black-box mode, PGD mode, and SMP mode). We compare the performance of the seven aggregation methods (FedAvg, NDC, RSA, RFA, Krum, and Multi-Krum) and the benchmark FedAvg*. We can see that some well-known robust methods are breached when the attack changed from the black-box mode to the PGD and SMP modes, and our method always maintains similar performance to FedAvg*.} \vspace{-4mm}
\label{fig:backdoor_exp}
\end{figure*}

\begin{figure*}[ht]
\begin{center}
\subfigcapskip = -8pt
\subfigure[Trigger \& Black-box]{
  \includegraphics[width=0.26\linewidth]{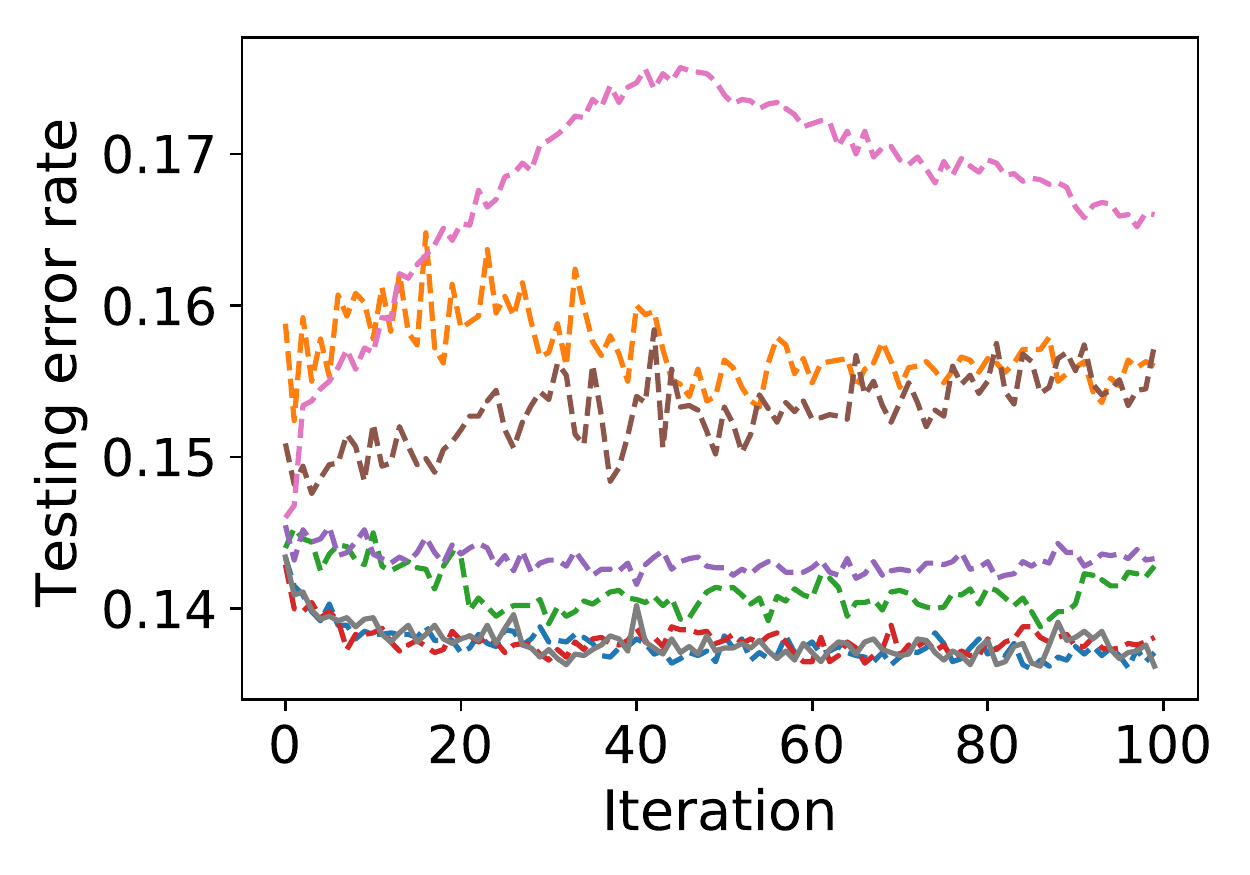}
}
\subfigure[Semantic \& Black-box]{
  \includegraphics[width=0.26\linewidth]{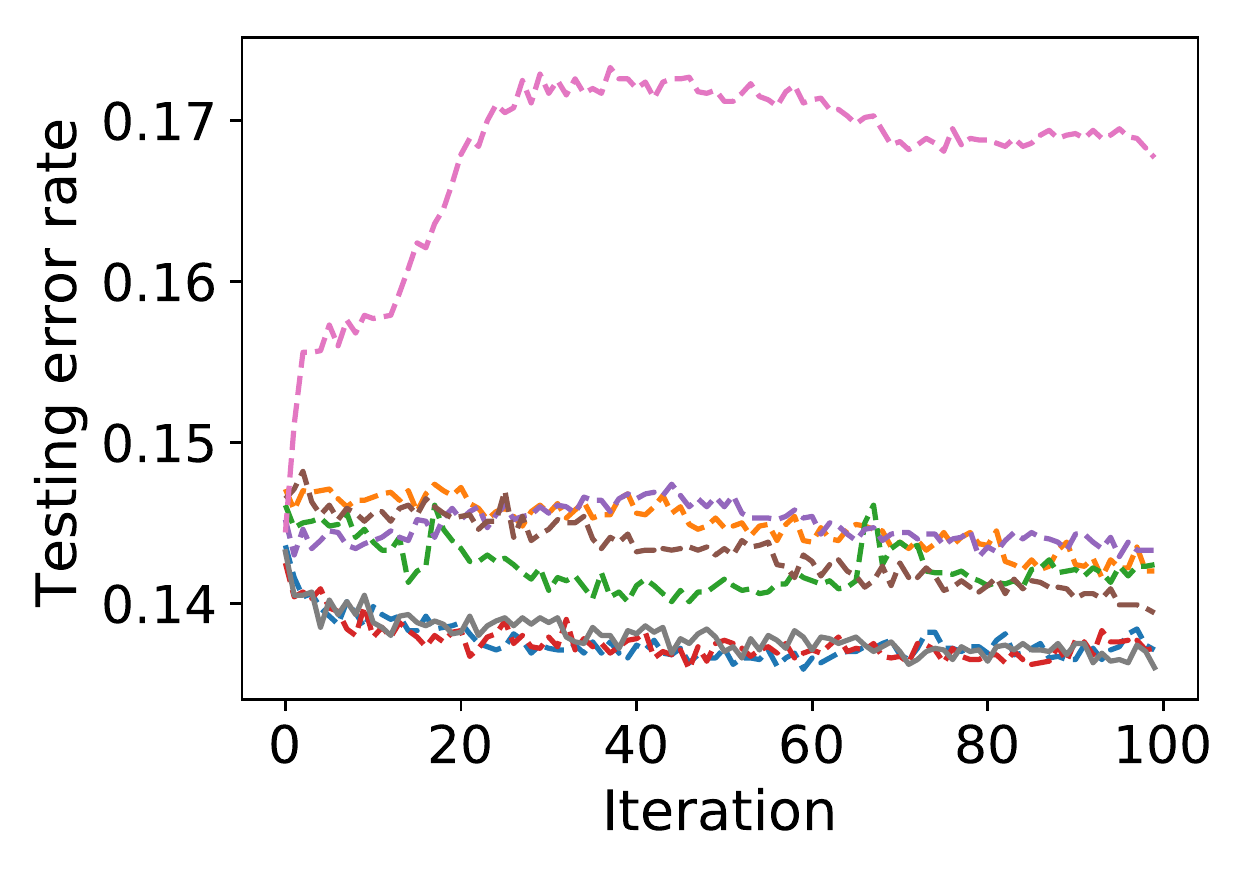}
}
\subfigure[Edge-case \& Black-box]{
\includegraphics[width=0.26\linewidth]{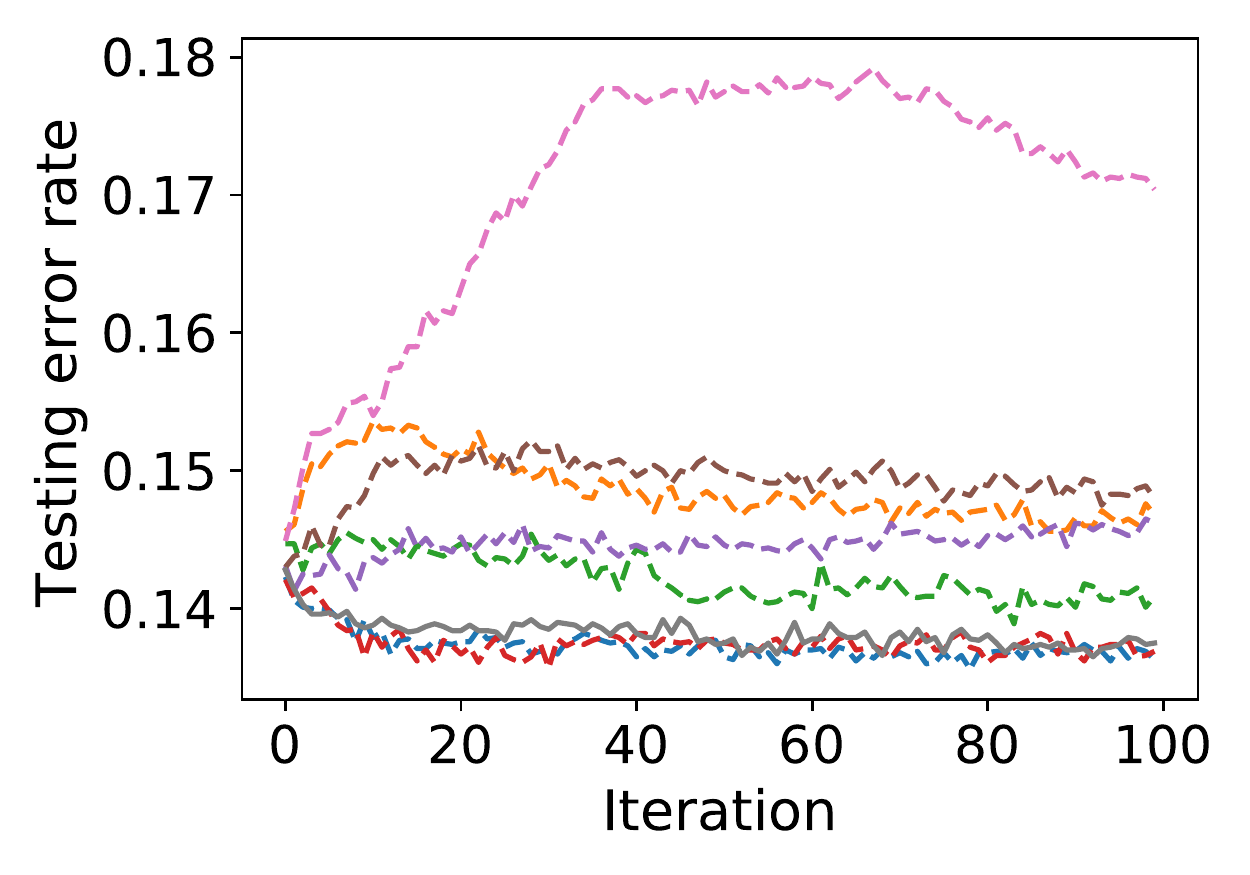}
}\vspace{-3mm}
\end{center}

\begin{center}
\subfigcapskip = -8pt
\subfigure[Trigger \& PGD]{
  \includegraphics[width=0.26\linewidth]{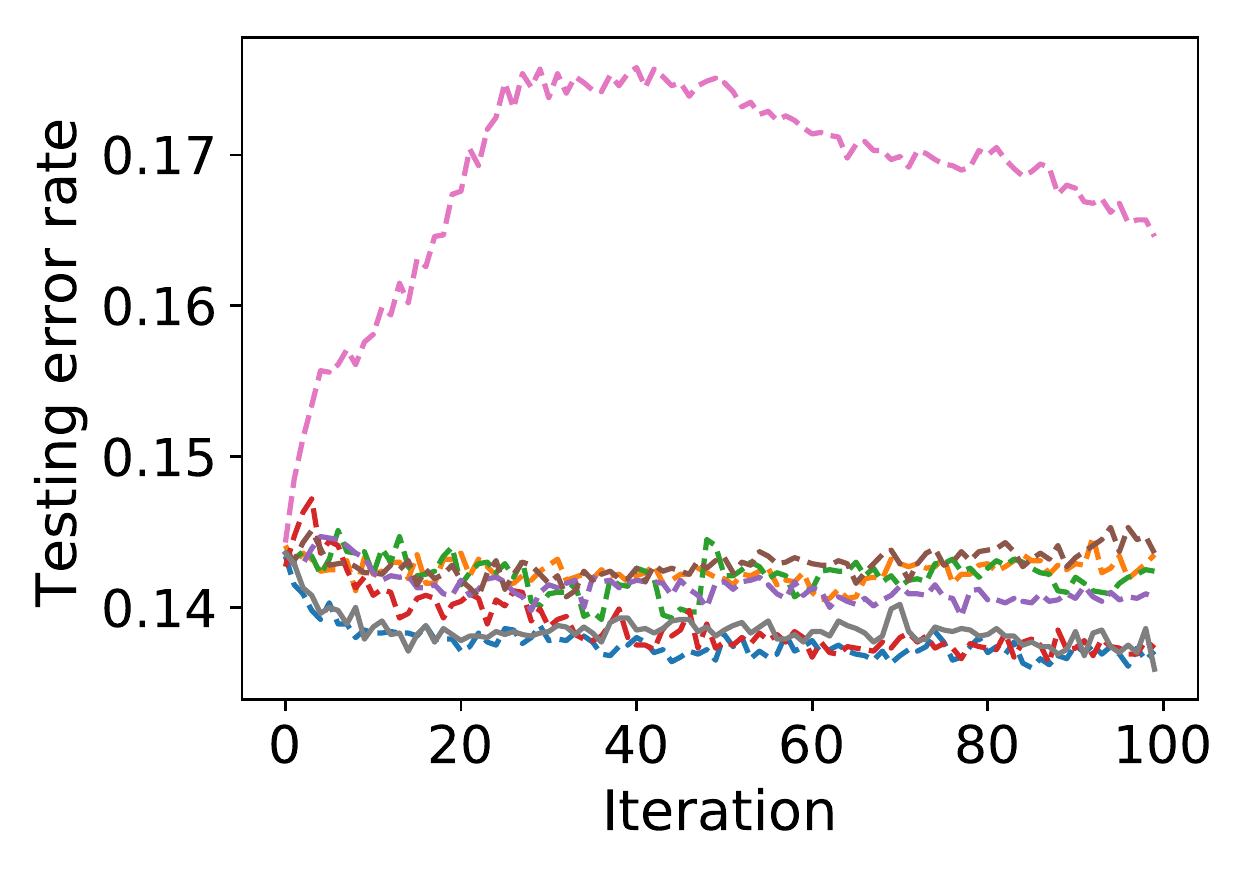}
}
\subfigure[Semantic \& PGD]{
  \includegraphics[width=0.26\linewidth]{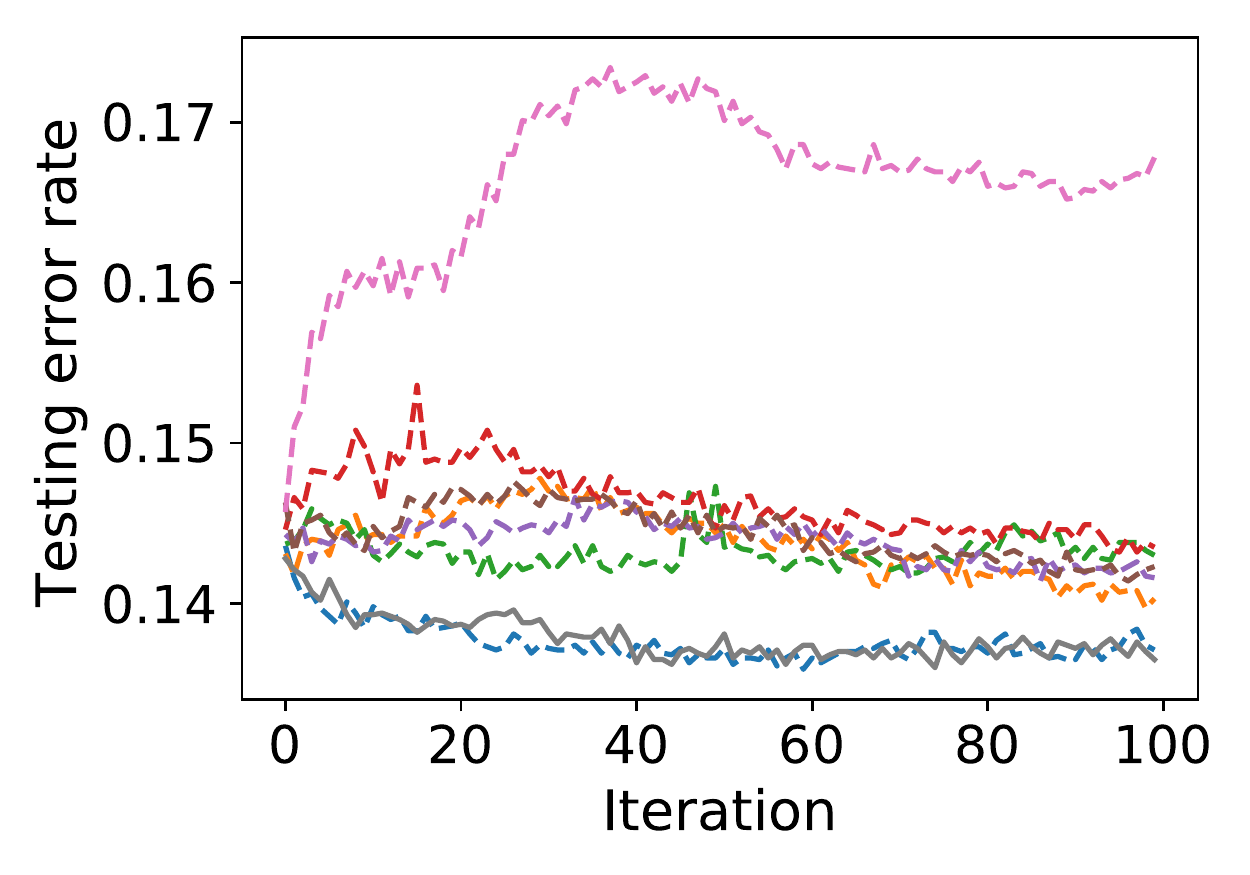}
}
\subfigure[Edge-case \& PGD]{
   \includegraphics[width=0.26\linewidth]{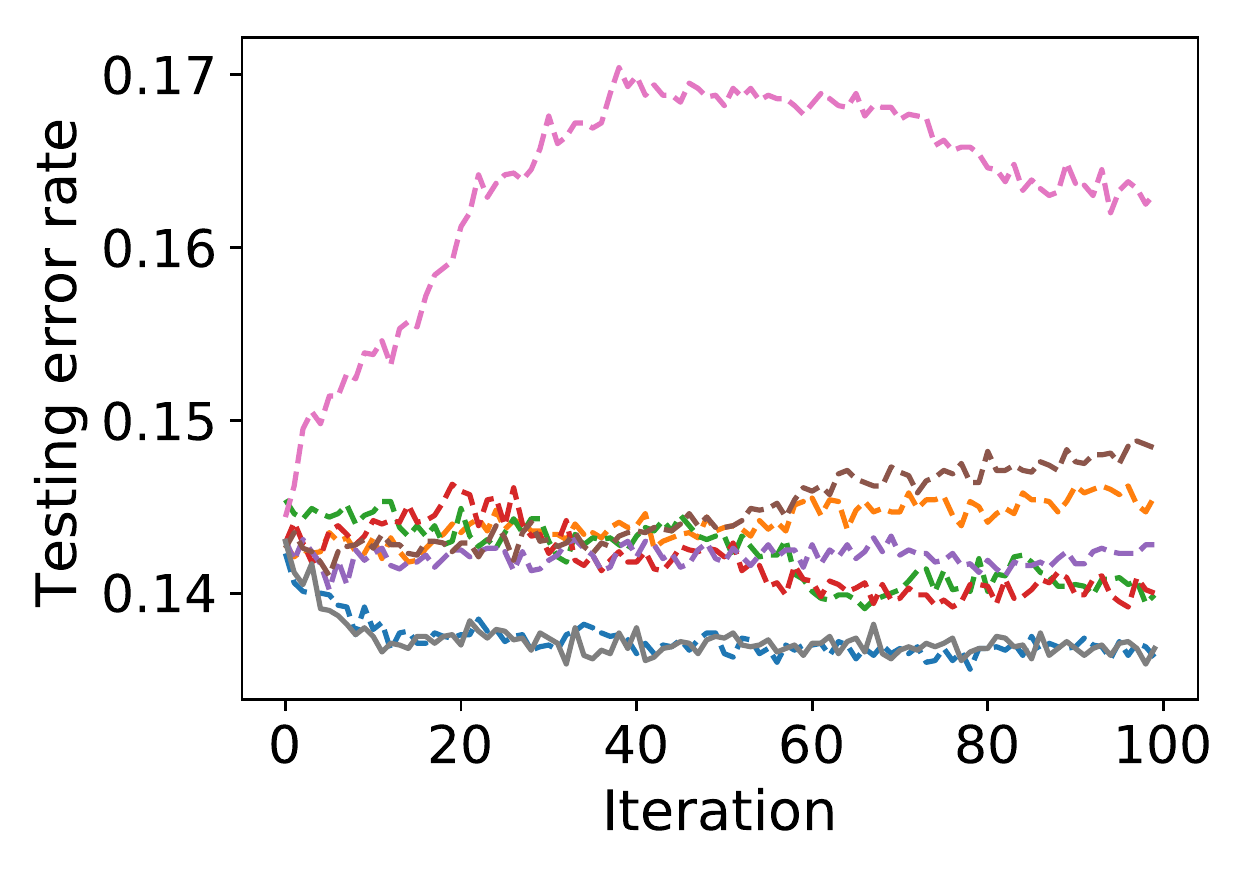}
}\vspace{-3mm}
\end{center}

\begin{center}
\subfigcapskip = -8pt
\subfigure[Trigger \& SMP]{
  \includegraphics[width=0.26\linewidth]{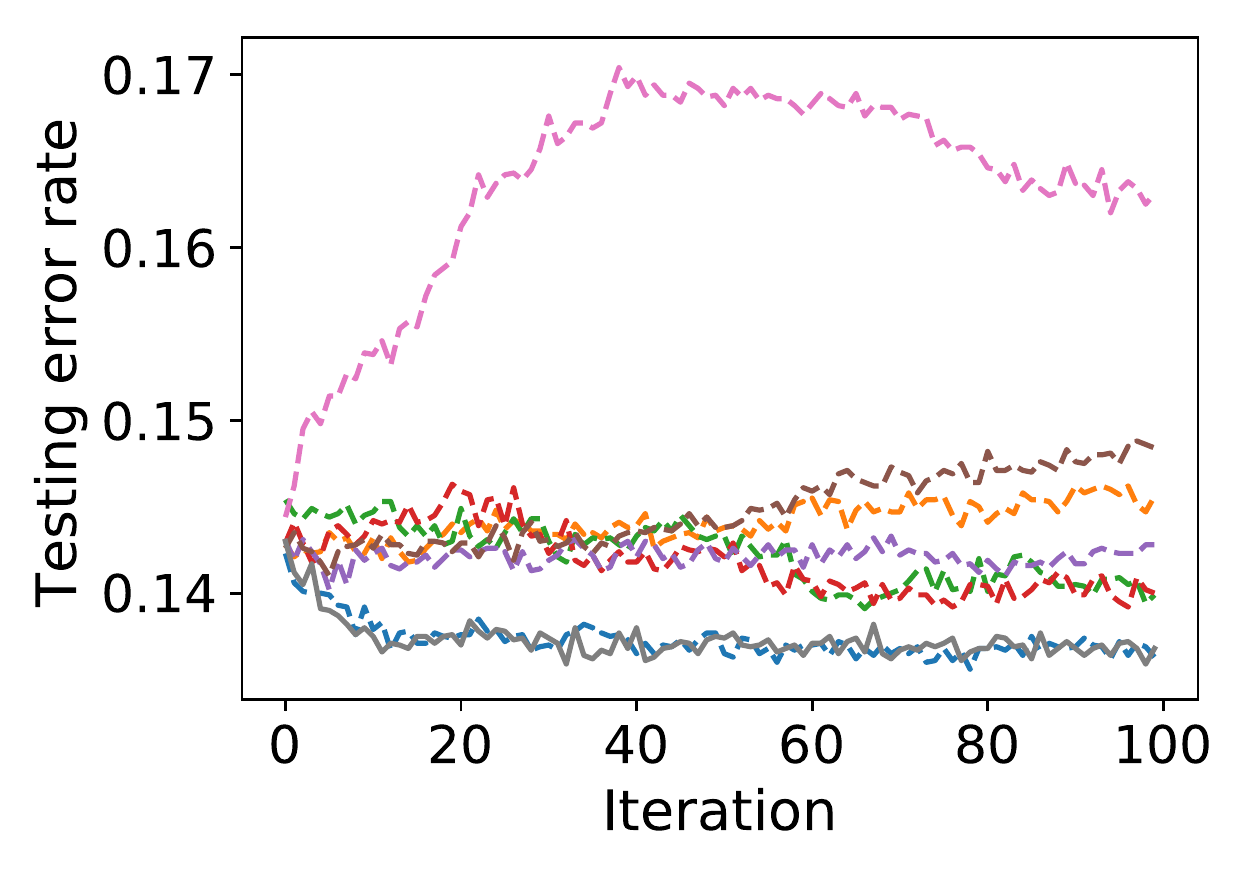}
}
\subfigure[Semantic \& SMP]{
  \includegraphics[width=0.26\linewidth]{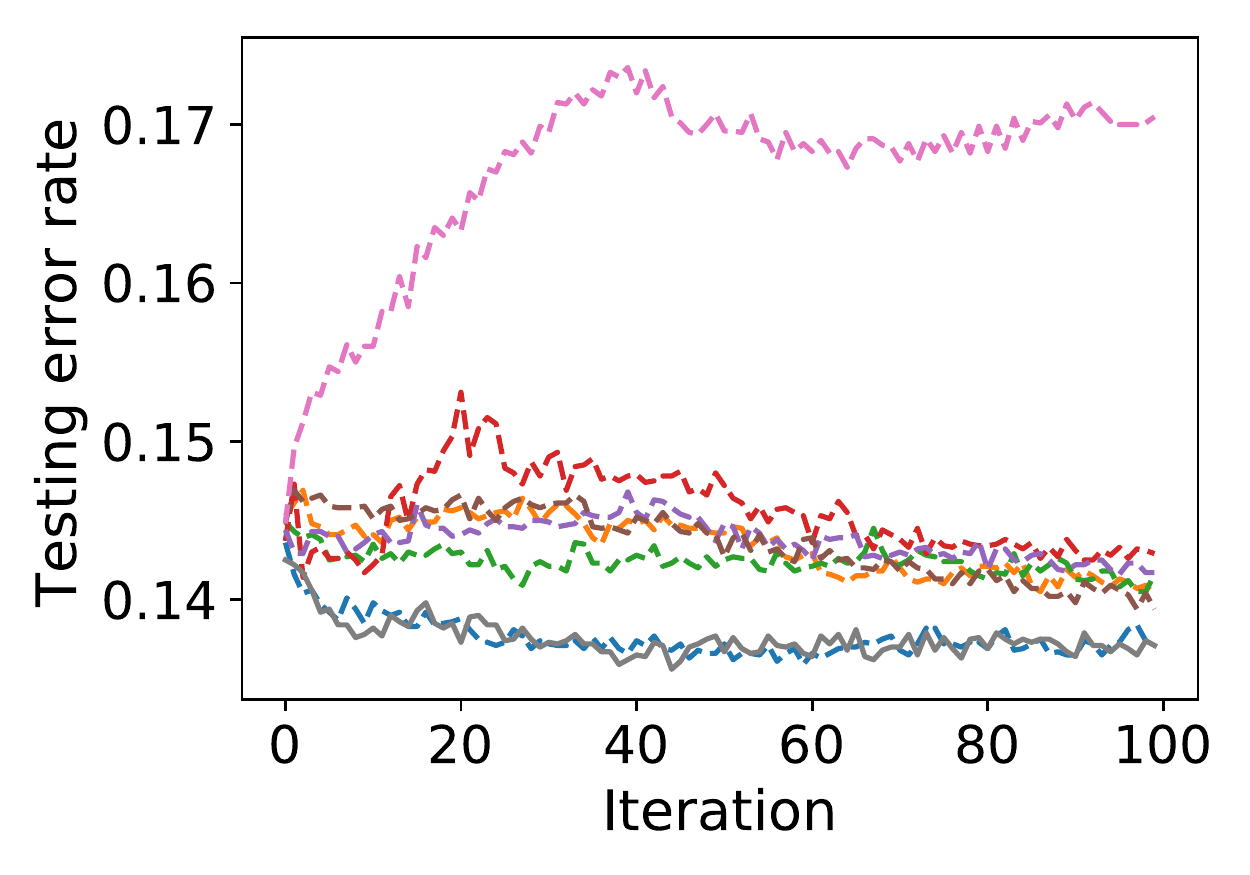}
}
\subfigure[Edge-case \& SMP]{
   \includegraphics[width=0.26\linewidth]{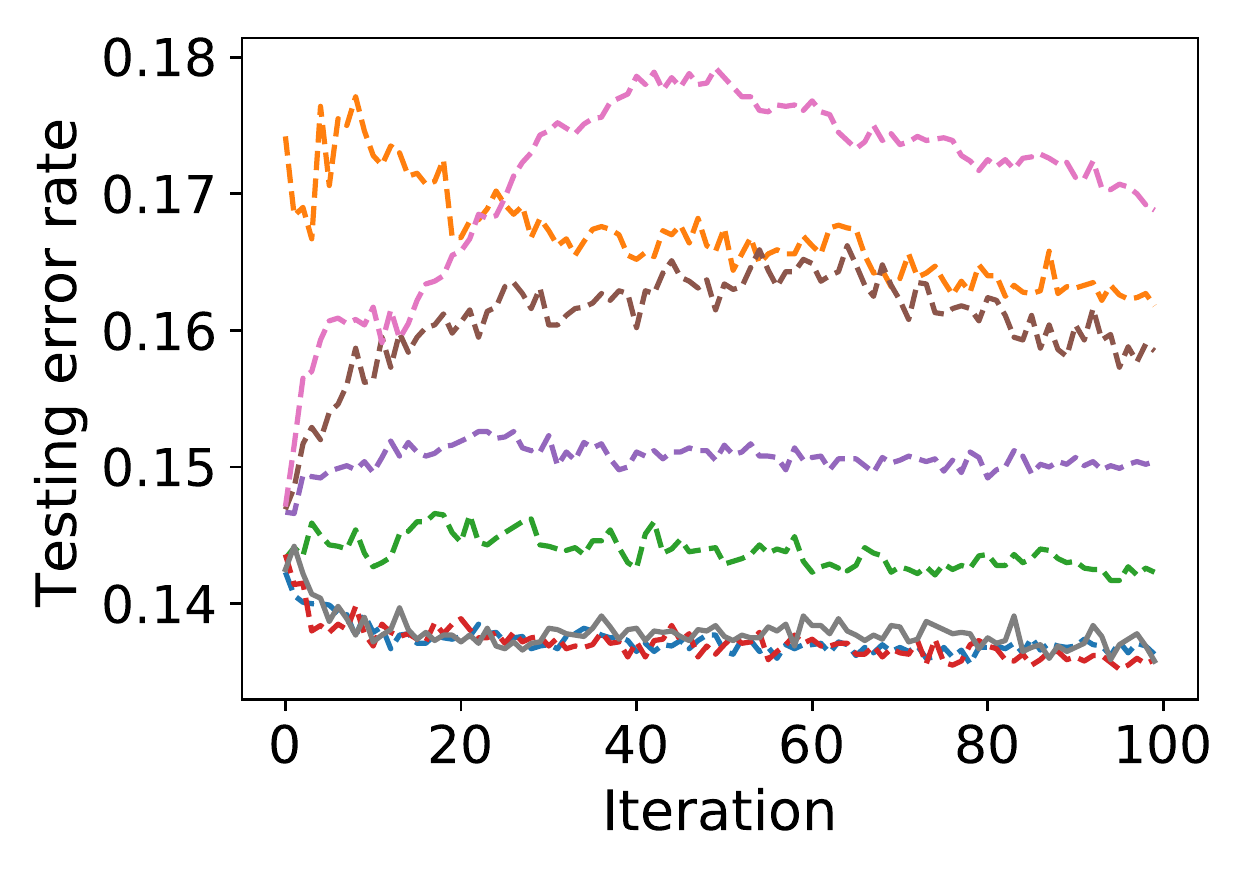}
}\vspace{-2mm}
\end{center}
\begin{center}
\includegraphics[width=0.8\linewidth]{tuli.pdf}
\end{center}\vspace{-3mm}
\caption{The attack success rate of three backdoor attacks (Trigger attack, Semantic attack, and Edge-case attack) on the main task under three modes (Black-box mode, PGD mode, and SMP mode). We compare the performance of the seven aggregation methods (FedAvg, NDC, RSA, RFA, Krum, and Multi-Krum) and the benchmark FedAvg*. We can see that in all backdoor attacks and modes, our aggregation method ensures a low testing error rate, which is comparable to FedAvg*. However, other methods, especially RSA, have a higher testing error rate than XMAM.} \vspace{-4mm}
\label{fig:backdoor_main}
\end{figure*}

Now, we explain the two special phenomena in Fig. \ref{fig:backdoor_exp}. The first one is that the attack success rate promotes about 30\% in Semantic backdoor attacks even when there is no malicious client. Why does it happen? We surmise that the data on green cars are close to the data on birds in the output space of the clean global model. To verify it, we separately set the target label of poisoned data from class 0 to class 9 and run 100 iterations to observe the performance of the global model on different backdoor tasks. The results show that the data of green cars is more likely to be classified as automobiles (class 1: 33.6\%), birds (class 2: 29\%), and frogs (class 6: 25.2\%) in a clean global model. Detailed results are in Fig. \ref{fig:analysis}. The second one is that Krum seems to have a more powerful capability than FedAvg* (FedAvg under no attacks) and XMAM in defending against Semantic attacks. However, as we explained in Figure 7, the global model has an innate accuracy of the backdoor task when there is no malicious client. Krum has a lower Attack Success Rate (ASR) because the global model trained by Krum has a lower innate accuracy of the backdoor task, which does not mean Krum is better than XMAM in defending against Semantic attacks. As in Krum, XMAM can also preclude malicious client models in each iteration with Semantic attacks. Therefore, we can conclude that Krum and XMAM both have the ability to evade Semantic attacks. However, XMAM will select more benign client models in each iteration when there are malicious clients so that XMAM has a good and similar performance to FedAvg* on the main task, but Krum is not good enough on the main task.

%---------------------------  Adaptive
\textbf{Performance on adaptive attacks:} The global model that uses Krum as the aggregation method has a worse convergence rate and a higher testing error rate, which is especially conspicuous when the data distribution is non-i.i.d. In this part, we further explain the weakness of Krum in defending against adaptive attack (Krum attack), which is designed by \cite{fang2020local} to challenge the robustness of Krum. We use Krum attack to test all aggregation methods on two datasets, CIFAR-10, and MNIST, Fig. \ref{fig:Adaptive} illustrates that Krum attack breakdowns the aggregation methods, Krum and Multi-Krum, and has a certain negative impact on the other methods except XMAM, which further demonstrates our method is more robust than Krum and Multi-Krum.

%To further verify the robustness of XMAM, we design a new adaptive attack, XMAM attack, according to the same framework of Krum attack. We find that however small the magnitude $\lambda$ is, the malicious local model updates will not be selected by XMAM. To find out why, we record the sum of Euclidean distance between the malicious local model update and the benign local model updates, and the sum of Euclidean distance between the malicious SLOU and the benign SLOUs. The results show that with the magnitude $\lambda$ decreasing, the malicious local model update and the benign local model updates are getting closer, but the malicious SLOU maintains almost the same distance to the benign SLOUs. Therefore, our method maintains the detection ability when the malicious local model updates become more and more hidden.

\begin{figure*}[ht]
\begin{center}
\subfigcapskip = -8pt
\subfigure[\tiny{$SLOU$ \& $\lambda=2^{-12}$}]{
  \includegraphics[width=0.16\linewidth]{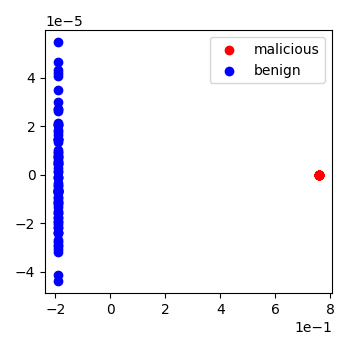}
}
\subfigure[\tiny{$SLOU$ \& $\lambda=2^{-14}$}]{
  \includegraphics[width=0.16\linewidth]{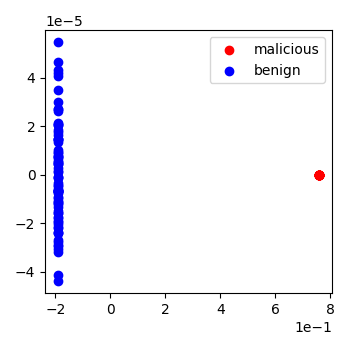}
}
\subfigure[\tiny{$SLOU$ \& $\lambda=2^{-16}$}]{
  \includegraphics[width=0.16\linewidth]{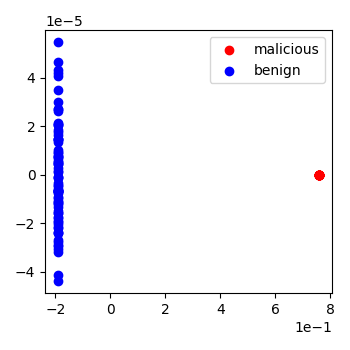}
}
\subfigure[\tiny{$SLOU$ \& $\lambda=2^{-18}$}]{
  \includegraphics[width=0.16\linewidth]{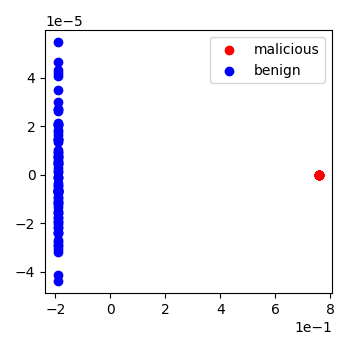}
}
\subfigure[\tiny{$SLOU$ \& $\lambda=2^{-20}$}]{
  \includegraphics[width=0.16\linewidth]{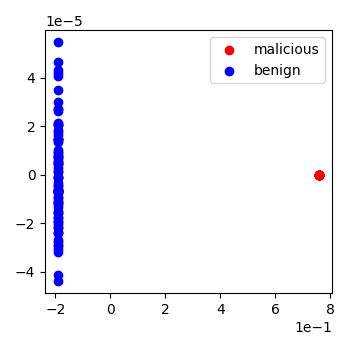}
}
\end{center}

\begin{center}
\subfigcapskip = -8pt
\subfigure[\tiny{update \& $\lambda=2^{-12}$}]{
  \includegraphics[width=0.16\linewidth]{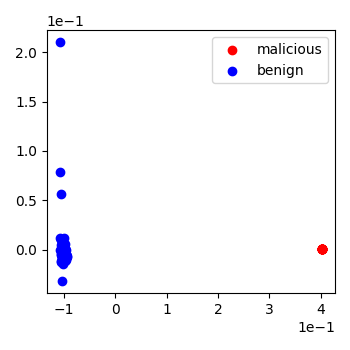}
}
\subfigure[\tiny{update \& $\lambda=2^{-14}$}]{
  \includegraphics[width=0.16\linewidth]{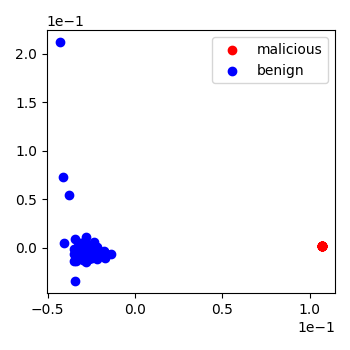}
}
\subfigure[\tiny{update \& $\lambda=2^{-16}$}]{
  \includegraphics[width=0.16\linewidth]{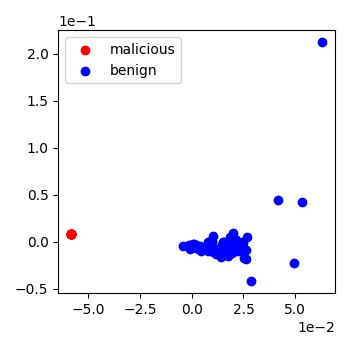}
}
\subfigure[\tiny{update \& $\lambda=2^{-18}$}]{
  \includegraphics[width=0.16\linewidth]{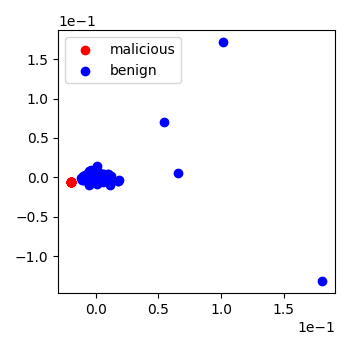}
}
\subfigure[\tiny{update \& $\lambda=2^{-20}$}]{
  \includegraphics[width=0.16\linewidth]{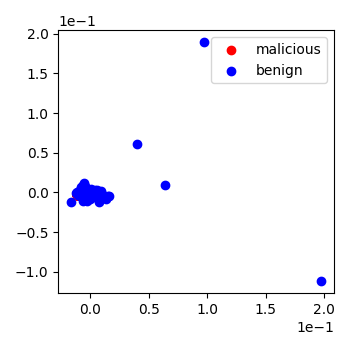}
}
\end{center}

% \begin{center}
% \includegraphics[width=0.8\linewidth]{tuli.pdf}
% \end{center}\vspace{-3mm}
\caption{The PCA scatter diagram of 100 client models' $SLOUs$ (top) and updates (bottom). We use CIFAR-10 and VGG9 in this experiment and malicious clients occupy 20 \%. } 
\label{fig:adap_eva}
\end{figure*}

\begin{figure}[ht]
\begin{center}  
\subfigcapskip=-6pt
\subfigure[CIFAR-10]{
    \includegraphics[width=0.3\linewidth]{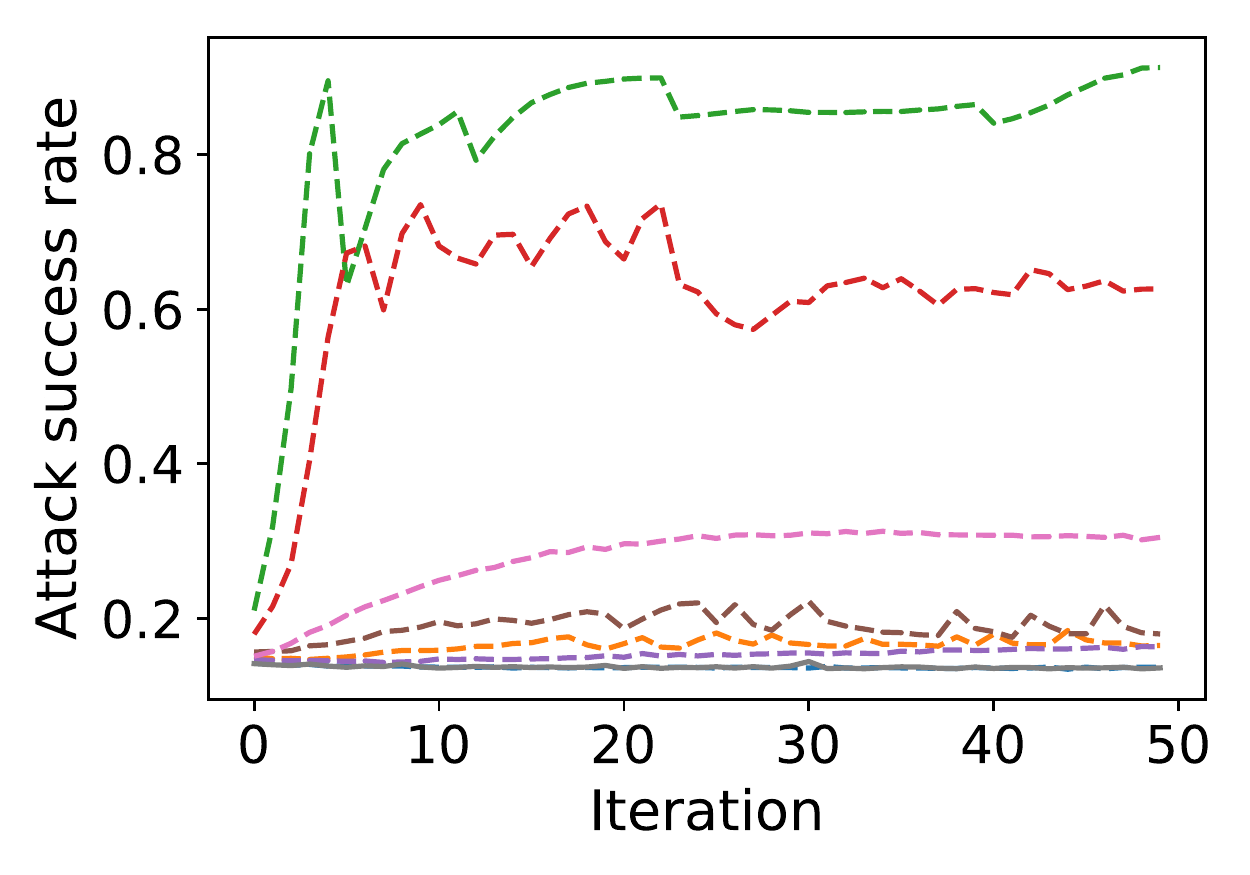}
}
\subfigure[MNIST]{
    \includegraphics[width=0.3\linewidth]{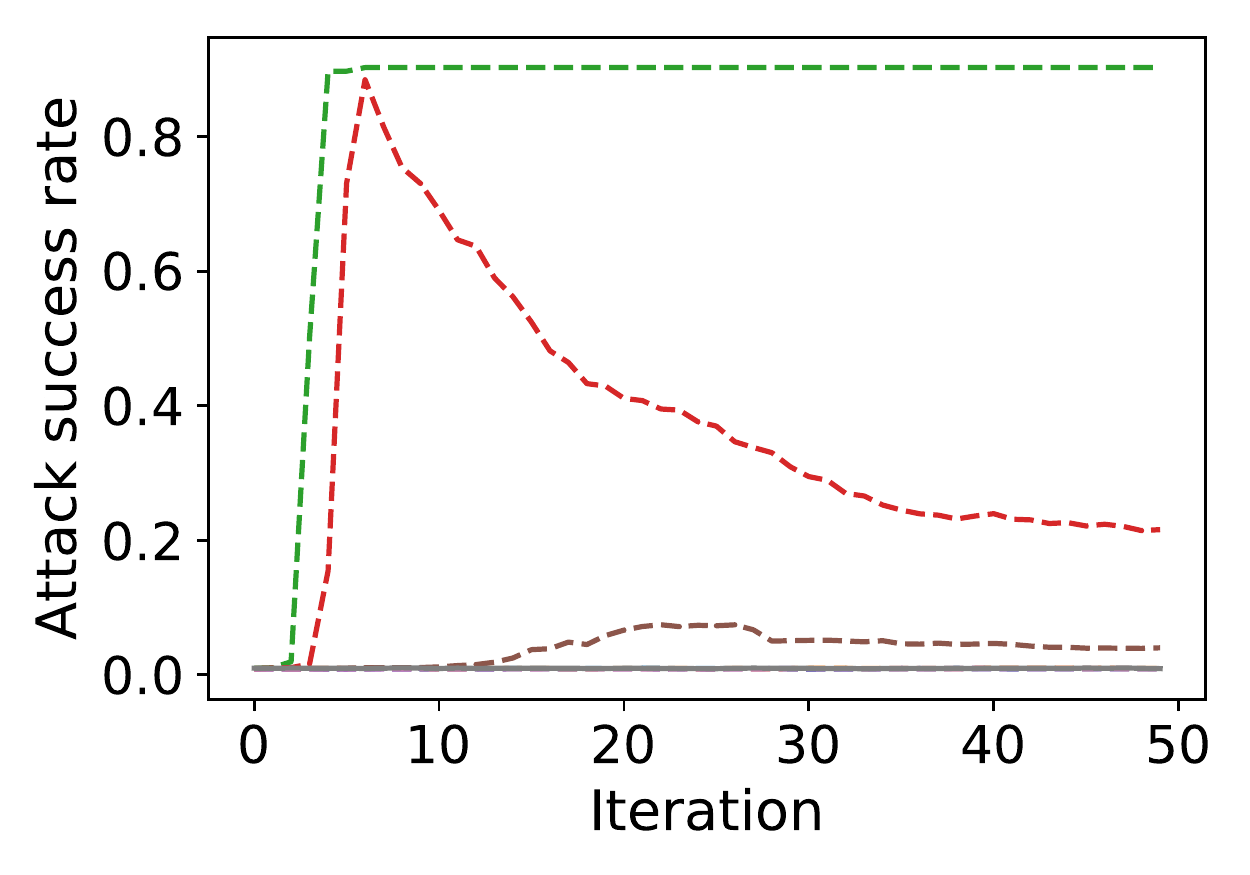}
}
\end{center}
\begin{center}
\includegraphics[width=0.4\linewidth]{tuli2.pdf}
\end{center}\vspace{-3mm}
\caption{The attack success rate of the Krum attack on two datasets (CIFAR-10 and MNIST). We can see from the results that Krum, and Multi-Krum seriously deteriorate and some other methods are slightly affected, but our method always has comparable performance to FedAvg*.} \vspace{-4mm}
\label{fig:Adaptive}
\end{figure}

To further verify the robustness of XMAM, we design a new adaptive attack, XMAM attack, according to the same framework as the Krum attack. We find that even $\lambda$ is very small, like $1e^{-10}$, XMAM will not select the malicious local model updates. To find out the root cause, we plot the scatter diagrams using Principal Component Analysis (PCA) algorithm to explain why XMAM does not fail in XMAM-adaptive attack, but Krum fails in Krum-adaptive attack. From Fig. \ref{fig:adap_eva} we can see that when the reverse magnitude $\lambda$ becomes smaller and smaller, the malicious clients' model updates become closer and closer to the benign client's model updates so that Krum can select the malicious clients' model updates. Nevertheless, with the reverse magnitude $\lambda$ becoming smaller and smaller, the distance between the malicious clients' $SLOU$ and the benign clients' $SLOU$ has no significant change. This experiment can illustrate XMAM's ability to defend against adaptive attacks.

3) \textit{Efficiency}: Now we compare the screening complexity of the proposed aggregation methods. The server using the aggregation method FedAvg does not need to screen received local model updates, so its screening complexity is $O(0)$. Krum and Multi-Krum compute the mutual distance of $\tau$ client local model updates. NDC and RSA clip and regularize $\tau$ local model updates, respectively. RFA finds the geometric center by considering $\tau$ client local model updates until it satisfies the defined condition. In a word, all the above methods need to consider the $\zeta$ parameters of the local model update. Unfortunately, the local model update in the deep learning model currently possesses millions of parameters (\textit{e.g.}, VGG-16 has 138M parameters). Instead of operating on the parameters of the local model update, our method considers the $M$ probabilities of the $Softmax$ layer's output. As far as we know, $M$ is far less than $\zeta$, which might be a millionfold.

For XMAM, the screening complexity covers two-parts. The first part is the time of training a random matrix to get the SLOUs. The second part is the time of clustering these SLOUs, which we set as $O(\tau^2 M R^*)$, where $R^*$ is the rounds needed by the clustering algorithm to find the cluster centers. Note that the time of training a random matrix could be negligible. Therefore, the ultimate screening complexity of XMAM is $O(\tau^2 M R^*)$. 

Table \ref{table:complexity} gives the concrete screening complexity for seven aggregation methods and the average screening time for ten arbitrary iterations. We run the procedure of screening on a computer with NVIDIA TITAN X GPU. The results show that compared with the same type of methods, Krum and Multi-Krum, XMAM improves the screening efficiency by tens of thousands of times. It is no surprise the XMAM reduces the dimension of the detection object from $\zeta$ to $M$.

\begin{table}
\centering
\caption{The screening complexity for seven aggregation methods. $\zeta$ denotes the number of local model update parameters, $M$ means the number of label classes and $\tau$ is the number of collected local model updates in each iteration. The screening time is recorded by experiments on dataset CIFAR-10 and network VGG9.} 
\label{table:complexity}
\scalebox{0.8}{
\begin{tabular}{|l|c|c|}
\hline
Method & Screening complexity & Screening time (second)\\
\hline\hline
FedAvg & $O(0)$ & 0 \\
Multi-Krum & $O(\tau^2 \zeta)$ & 116.98 \\
Krum & $O(\tau^2 \zeta)$ & 109.19 \\
RFA & $O(\tau \zeta R^*)$ & 39.93 \\
RSA & $O(\tau \zeta)$ & 0.78 \\
NDC & $O(\tau \zeta)$ & 0.34 \\
XMAM & $O(\tau^2 M R^*)$ & 0.0079 \\
\hline
\end{tabular}
}
\end{table}

\section{Discussions and limitations}

\textbf{Discussion:} Huang \cite{huang2019neuroninspect} uses the interpretability technique to design a heatmap to explain the DNNs output, facilitating a more accurate detection for the poisoned model. However, their method requires a clean dataset encompassing all classes. Different from their method, our scheme only needs a generated random matrix, and it directly uses the probability distribution of the $Softmax$ layer's output to judge the quality of a model. Kolouri\cite{kolouri2020universal} feeds a group of Universal Litmus Patterns (ULPs) through a model and pools the logit layer's output to classify it as poisoned or clean. Nevertheless, optimizing the classifier and ULPs requires hundreds of pre-trained clean and poisoned models, which is impractical in FL because the server has no dataset. Huang\cite{huang2020one} proposes One-Pixel Signature for backdoor detection. As in \cite{kolouri2020universal}, One-Pixel Signature also demands pre-trained clean and poisoned models. 

\textbf{Limitations:} As we know, the data distribution in FL is in a non-i.i.d. scenario, and our experimental datasets are also distributed in a non-i.i.d. way. In the early stages of training, the local model updates submitted by the benign clients are more heterogeneous than the local model updates in the middle and late stages of training. Therefore, our method is not guaranteed to be effective at the early stages of training. Note that in the middle and late stages of training, our method will occasionally fail to preclude all malicious local model updates in a certain round, but this does not mean that our defense is ineffective because backdoor attacks require multiple rounds to succeed. 

\begin{figure}[ht]
\begin{center}  
\subfigcapskip=-6pt
\subfigure[Black-box]{
    \includegraphics[width=0.3\linewidth]{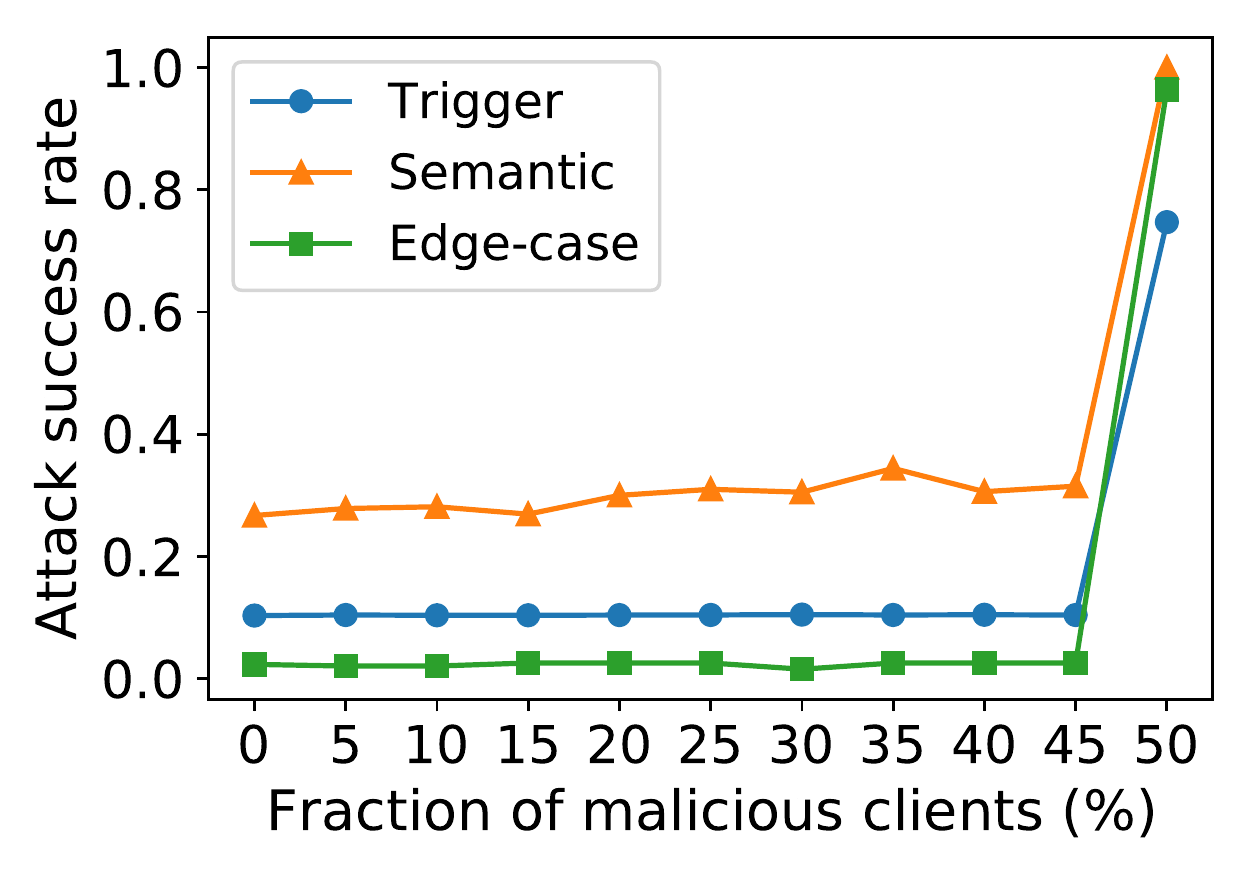}
}\vspace{-1mm}
\subfigure[PGD]{
    \includegraphics[width=0.3\linewidth]{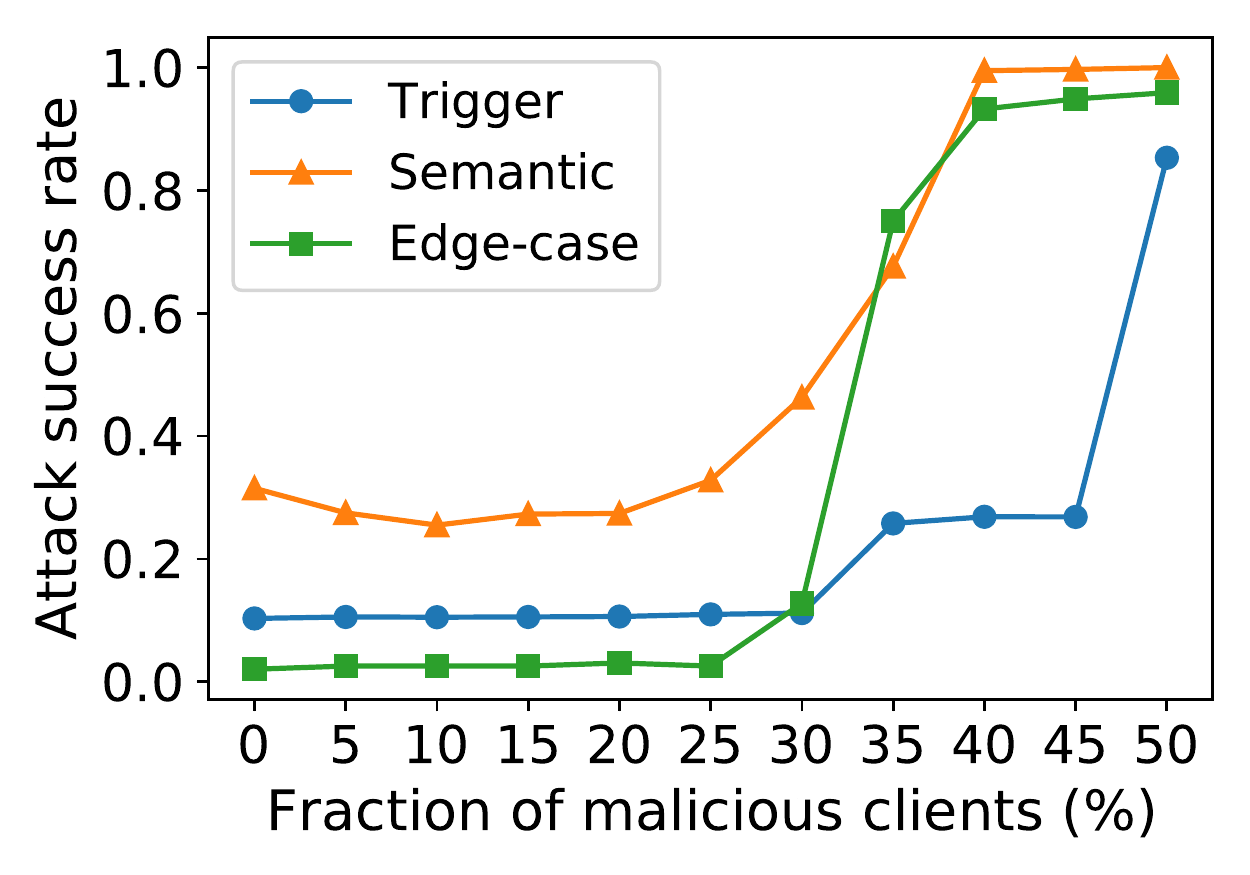}
}
\caption{The attack success rate of three backdoor attacks under different proportions of malicious clients. We can see that our defense can tolerate 45\% malicious clients under black-box mode can about 30\% malicious clients under PGD mode.} \vspace{-4mm}
\label{fig:backdoorratio}
\end{center}
\end{figure}

In the above experiments, we demonstrate that our method is the most robust aggregation method compared with the other six under a certain proportion (20\%) of malicious clients. Then, we conduct an experiment to observe our method's tolerance in different proportions of malicious clients. The results (Fig. \ref{fig:backdoorratio}) show that our method can tolerate at least 45\% of malicious clients in the black-box mode and about 30\% of malicious clients in the PGD mode. However, we must admit that the results of our defense are still not enough, and we will find ways to improve XMAM's tolerance to a larger number of malicious clients in future work.

%------------------------------------------------------------------------

\section{Related works}

Poisoning attacks refer to destroying the training result of machine learning. A typical poisoning attack is data poisoning attacks\cite{biggio2012poisoning,chen2017targeted,fang2018poisoning,gu2017badnets,jagielski2018manipulating,li2016data,munoz2017towards,nelson2008exploiting,rubinstein2009antidote,shafahi2018poison,suciu2018does,wang2019attacking,xiao2015feature,yang2017fake,fang2021data}. Recently, multiple poisoning attacks \cite{bagdasaryan2020backdoor,xie2019dba,wang2020attack} have threatened FL. In addition to data poisoning attacks, FL also suffers from local model poisoning attacks\cite{fang2020local,li2019rsa,baruch2019little,xie2020fall,he2020byzantine,bagdasaryan2020backdoor,wang2020attack,bhagoji2019analyzing}, which are more potent than the former. Furthermore, from the perspective of the attack's purpose, these poisoning attacks can be divided into untargeted attacks\cite{fang2020local,li2019rsa,baruch2019little,xie2020fall,he2020byzantine}, which aim to deteriorate the global model, and backdoor (targeted) attacks\cite{bagdasaryan2020backdoor,bhagoji2019analyzing,xie2019dba,wang2020attack}, which aim to induce the global model to make some attacker-chosen mistake in certain input without deteriorating the global model. The latter is more threatening to FL owing to its hidden characteristic. 

In central learning, all data is used to train a model jointly, and it is assumed that the model is transparent to the attacker. Therefore, in the poisoning-based backdoor attacks \cite{gu2017badnets,chen2017targeted,liao2018backdoor,liu2017trojaning,zhao2020clean,bagdasaryan2020backdoor}, the attacker can optimize the trigger in the poisoned data according to the performance of the model on them. In FL, the attacker does not know what the global model of the current iteration will be, so it is impossible to optimize the trigger in the poisoned data based on the performance of the global model on them. In addition, apart from the poisoning-based backdoor attacks, when the data can not be tampered with, the backdoor attacker in the central learning can also achieve the backdoor function by directly perturbing the model weight, which is the so-called \textit{Targeted Weight Perturbation} (TWP) \cite{dumford2020backdooring}. Similarly, this is difficult to achieve in FL because the backdoor attacker in FL can only control its own model, and can not directly tamper with the global model. Currently, there are three types of backdoor attacks in FL: Trigger backdoor \cite{gu2017badnets}, Semantic backdoor \cite{li2020backdoor}, and Edge-case backdoor \cite{wang2020attack}. All of them belong to the poisoning-based backdoor attacks. 

There are three types of defense approaches proposed in centralized learning \cite{li2020backdoor}: i) \textit{trigger-backdoor mismatch}, ii) \textit{trigger elimination}, and iii) \textit{bcakdoor elimination}. The \textit{trigger-backdoor mismatch} based defenses \cite{liu2017neural,doan2020februus,udeshi2019model,qiu2021deepsweep} try to modify or reverse the trigger in the poisoned data so that the modified poisoned data can not match the hidden backdoor in the model to prevent backdoor activation. The \textit{trigger elimination} based defenses \cite{gao2019strip,subedar2019deep,du2019robust,javaheripi2020cleann} attempt to distinguish between poisoned data and benign data, and preclude the poisoned data before training. Both of the above two types of defenses require direct contact with client data, which is impractical in FL. The \textit{bcakdoor elimination} based defenses \cite{wang2019neural,chen2019deepinspect,huang2019neuroninspect,xu2019detecting,huang2020one} try to prune the neurons in the model that can be activated by the trigger, or refuse to deploy the infected model through detection. This type of defense requires the server having a large batch of data or plenty of models that have been labeled as infected or uninfected to train a meta-classifier, which is difficult to obtain in FL. 

The existing defenses in FL can be roughly divided into three categories: (i) \textit{limiting the update}, (ii) \textit{finding the ``center''}, and (iii) \textit{detecting and precluding}. The first category is \textit{limiting the update} \cite{sun2019can,li2019rsa}. The core of this type of defense is to punish the local model update that has a large norm or regularize all the local model updates to a small norm. The second category is \textit{finding the ``center''} \cite{pillutla2019robust}. The core of this type of defense is to exploit the local model updates submitted by clients to find a compromised update, which is the ``center'' of the local model updates, to update the global model. The third category is \textit{detecting and precluding} \cite{blanchard2017machine}. The core of this type of defense is to detect the malicious local model update and preclude it.

\section{Conclusion} 

We proposed a new aggregation method, XMAM, to reveal backdoor attacks that use hiding techniques in FL. Unlike the existing aggregation methods that focus on the parameters of local model updates, we focus on the SLOUs. The extensive evaluations show that our method can effectively distinguish malicious local model updates from benign ones without any training dataset in the server. Specifically, when other methods fail to defend against the backdoor attacks at no more than 20\% malicious clients, our method can tolerate 45\% malicious clients in black-box modes and about 30\% in PGD mode. Besides, our XMAM is resilient to adaptive attacks even when there are 40\% malicious clients. Finally, we analyze our method's screening complexity and compare the real screening time with other methods. The results show that XMAM is about 10-10000 times faster than the existing methods. Interesting future work includes 1) considering a more effective matrix that can be optimized to adapt to our detection and 2) designing a stronger detection method that not only focuses on the $Softmax$ layer's output.

\section*{Acknowledgment}
We thank the anonymous reviewers for their constructive comments.

%-\frac{2l^{4}\tan^{2}\alpha}{\cos\alpha})$ be $g(\alpha)$. \\
%Let $h=2l,2.5l,3l,3.5l,4l$ respectively, if there exists $\alpha_0$, $\alpha_1$ and $\alpha_2$ satisfying:\\
% \text{~~~~~~~~}$g(\alpha_1)'=0$, \\
% \text{~~~~~~~~}$g(\alpha_0)'<0$,\\
% \text{~~~~~~~~}and $g(\alpha_2)'>0$ \\
% where $\alpha_0$ is minor smaller than $\alpha_1$,$\alpha_2$ is minor larger than $\alpha_1$,
% then the expected $\alpha$ can be derived. Unfortunately, when  $g(\alpha_1)'=0$, $\alpha\notin\lbrack0,\frac{\pi}{2}\rbrack$.
% We substitute the series values of h to $g(\alpha)$, then
% achieve the minimal values of $g(\alpha)$ and their corresponding values of $\alpha$. The average value of $\alpha$ is 1.05.
% $\tan1.05=1.74$, so the depth of hole information announcement is:\\
% \text{~~~~~~~~}$1.74*l=1.74*\frac{L}{2}=0.87L$,\\
% Note  here $l$ is approximately represented by  $\frac{L}{2}$.
%
%%%%%%%%%%%%%%%%%
%%%%%%%%%%%%%%comment ends 12/25/14

%%%%%%%%%%%%%%%

\bibliographystyle{unsrtnat}
\bibliography{references}

\appendix
\subsection{Feasibility analysis for XMAM} \label{app:proof}

If malicious local model update comes to XMAM, and it turns $a_{22}$ into $a_{22}+\lambda_{1}$ and $b $ into $b+\lambda_{2}(\lambda_{1}>0,\lambda_{2}>0)$ :\newline

\begin{center}
 $Relu_{m}=Relu\Bigg(\begin{bmatrix}A_{1,1}^{\prime} & \dots & A_{1,n-2}^{\prime} \\ \vdots &\ddots &\vdots \\ A_{n-2,1}^{\prime} & \dots & A_{n-2,n-2}^{\prime} \end{bmatrix}\Bigg)$\newline
 
 $(A_{p,q}^{\prime}=A_{p,q}+\lambda_{1} i_{p+1,q+1}+b+\lambda_{2})$\newline
 
 $Pool_{m}=maxpooling(Relu_{m})$\newline
 
 $output=\begin{bmatrix}P_{1,1}^{\prime} & \dots & P_{1,n-4}^{\prime} \\ \vdots &\ddots &\vdots \\ P_{n-4,1}^{\prime} & \dots & P_{n-4,n-4}^{\prime} \end{bmatrix}\times
\begin{bmatrix}s_{1}\\ \vdots \\ s_{n-4} \end{bmatrix}
+\begin{bmatrix}\hat{b_{1}}\\ \vdots \\ \hat{b_{n-4}} \end{bmatrix}
=$\newline
$\begin{bmatrix}\sum{P_{1,i}^{\prime}s_{i}}+\hat{b_{1}}\\ \vdots \\ \sum{P_{n-4,i}^{\prime}s_{i}}+\hat{b_{n-4}} \end{bmatrix}$
$(P_{p,q}^{\prime}=P_{p,q}+\lambda_{1} i_{p+1,q+1}+b+\lambda_{2})$

$SLOU_{m}=softmax(output_{m})$\newline

$=\begin{bmatrix}slou_{a,1}\\ \vdots \\ slou_{a,n-2} \end{bmatrix}$\newline
\end{center}

Only when $i_{2,2}=i_{2,3}=\dots=i_{n-1,n-1}$, $SLOU_{m}=SLOU$, which is of low probability. Otherwise, $SLOU_{m}\neq SLOU$.
  
%  $\begin{bmatrix}a_{1,1} & a_{1,2} & a_{1,3} \\ a_{2,1} &a_{2,2} &a_{2,3} \\ a_{3,1} & a_{3,2} & a_{3,3} \end{bmatrix}\xrightarrow{malicious_update}   \begin{bmatrix}a_{1,1} & a_{1,2} & a_{1,3} \\ a_{2,1} &a_{2,2}+\lambda &a_{2,3} \\ a_{3,1} & a_{3,2} & a_{3,3} \end{bmatrix}$\newline
% $output_{a}=Relu(\begin{bmatrix}\sum{s_{i}}(\sum{a}+\lambda)+b_{1}\\ \vdots \\ \sum{s_{i}}(\sum{a}+\lambda) +b_{n-2} \end{bmatrix})=Relu(\begin{bmatrix}\sum{s_{i}}\sum{a}+\lambda \sum{s_{i}}+b_{1}\\ \vdots \\ \sum{s_{i}}\sum{a}+\lambda \sum{s_{i}}+b_{n-2} \end{bmatrix})$\newline

% $SLOU_{a}=softmax(output_{a})$\newline

% $=\begin{bmatrix}SLOU{a,1}\\ \vdots \\ SLOU{a,n-2} \end{bmatrix}$\newline
% \end{center}

%   Since there exists $k\in [1,n-2]$ such that $\sum{s_{i}}\sum{a}+b_{k}<0 $ or $\sum{s_{i}}\sum{a}+\lambda \sum{s_{i}}+b{k}<0$, 
%   if all the $k$ satisfy $\sum{s_{i}}\sum{a}+b_{k} < 0$ or $\sum{s_{i}}\sum{a}+\lambda \sum{s_{i}}+b{k}<0$, then $output=\begin{bmatrix}1\\ \vdots \\ 1 \end{bmatrix}$, which makes no sense. If some $k$s satisfy the conditions above, then $output_{a}-output\neq \begin{bmatrix}\lambda \sum{a}\\ \vdots \\ \lambda \sum{a} \end{bmatrix}$ ,which leads to  $SLOU_{a}\neq SLOU$.
  
The same procedure may be easily adapted to obtain the result for any other $a_{i,i}$.
  
And if malicious update turns $s_{m} $ into $s_{m}+\lambda(\lambda>0)$:\newline

\begin{center}
$\begin{bmatrix}s_{1}\\  \vdots \\ s_{m} \\ \vdots \\ s_{n-4} \end{bmatrix}\xrightarrow{malicious_update}  
\begin{bmatrix}s_{1}\\  \vdots\\ s_{m}+\lambda \\ \vdots \\ s_{n-4}
\end{bmatrix}$\newline
$output_{s}=\begin{bmatrix}\sum{P_{1,i}s_{i}}+\lambda P_{1,m}+\hat{b_{1}}\\ \vdots \\ \sum{P_{n-4,i}s_{i}}+\lambda P_{n-4,m}+\hat{b_{n-4}} \end{bmatrix}$\newline

$SLOU_{s}=softmax(output_{s})$\newline

$=\begin{bmatrix}slou_{s,1}\\ \vdots \\ slou_{s,n-2} \end{bmatrix}$\newline
\end{center}

Only when $P_{1,m}=P_{2,m}=\dots=P_{n-4,m}$, $SLOU_{s}=SLOU$, which is of low probability. Otherwise, $SLOU_{s}\neq SLOU$.

  if malicious update turns $\hat{b_{m}}$ into 
$\hat{b_{m}}+\lambda(\lambda>0)$:

\begin{center}
$\begin{bmatrix}\hat{b_{1}}\\ \vdots \\ b_{m} \\ \vdots \hat{b_{n-4}} \end{bmatrix}\xrightarrow{malicious_update}  
\begin{bmatrix}\hat{b_{1}}\\ \vdots \\ b_{m}+\lambda \\ \vdots \hat{b_{n-4}} \end{bmatrix}$\newline
$output_{\hat{b}}=\begin{bmatrix}\sum{P_{1,i}s_{i}}+\hat{b_{1}}\\ \vdots \\ \sum{P_{m,i}s_{i}}+\hat{b_{m}}+\lambda \\ 
\vdots \\ \sum{P_{n-4,i}s_{i}}+\hat{b_{n-4}} \end{bmatrix}$\newline

$SLOU_{\hat{b}}=softmax(output_{\hat{b}})$\newline

$=\begin{bmatrix}slou_{\hat{b},1}\\ \vdots \\ slou_{\hat{b},n-2} \end{bmatrix}$\newline
\end{center}

  It is obvious that $SLOU_{\hat{b}}\neq SLOU$.
  
\subsection{Explanations of notations}
See Table \ref{tab:notations} for explanations of notations.

\begin{table}
\begin{center}
\caption{Explanations of notations.}
\label{tab:notations}
\scalebox{0.6}{
\begin{tabular}{|l|c|}
\hline
notations & explanations \\
\hline\hline
$w_i^t$ & the $i^{th}$ client's updated model at $t^{th}$ iteration\\
$\hat{w}^t$ & the $i^{th}$ malicious client's local model at $t^{th}$ iteration\\
$w_{\textit{g}}^t $ & the global model at $t^{th}$ iteration \\
$u_i^t$ & the $i^{th}$ client's local model update at $t^{th}$ iteration\\
$\hat{u}^t$ & the $i^{th}$ malicious client's local model update at $t^{th}$ iteration\\
$u_{\textit{g}}^t $ & the global model's update at $t^{th}$ iteration \\ 
${u}_i^c$ & the local model update after clipping in NDC \\ 
${u}^{\star}$ & the selected model update in Krum \\ 
$M$ & the number of classes of dataset \\
$N$ & the total number of participating clients \\
$\tau$ & the number of client models the server collect in each iterations\\
$\tau^{'}$ & the number of remained models after screening by XMAM \\
$f$ & the number of Byzantine clients in $\tau$ clients at each iteration\\
$x_{i,j}$ & the $j^{th}$ training data of $i^{th}$ client\\
$\mathcal{D}_i$ & the dataset of $i^{th}$ client\\
$|\mathcal{D}_i|$ & the number of the $i^{th}$ client's training data\\ 
$\mathcal{D}_{mat}$ & the generated matrix \\
$\mathcal{D}_{p}$ & the poisoned data of client's training data \\
$\mathcal{D}_{c}$ & the clean data of client's training data \\
${\eta}_{\text{g}}$ & the global learning rate \\
$\delta$ & the clipping parameter of NDC\\
$\beta_r$ & the learning rate of RSA\\
$p_i$ & the weight according to the number of data of $i^{th}$ client\\
$q_i^r$ & the intermediate result of $i^{th}$ client in RFA\\
$v$ & smoothing factor of RFA\\
$z^r$ & the geometric median point in $r^{th}$ round in RFA \\
$R$ & the maximum number of rounds for per iteration in RFA \\
$\mu$ & the fault tolerance threshold in RFA \\
$\epsilon$ & the PGD parameter\\
${\rho}_{1}, {\rho}_{2}$ & the hyper-parameters of adversarial objective in SMP mode \\
$SLOU$ & the Softmax layer's output of model whose parameter is local model update \\
$Clusters$ & the clustering result by using HDBSCAN algorithm \\
${Cluster}_{major}$ & the set of nodes' id in the largest cluster \\
$\lambda$ & the magnitude that maximizes the bounded attack effect in adaptive attacks \\
${u}^{'}$ & the malicious local model update \\
${s}$ & the output of $Sign()$ of the global model update \\
$\chi$ & the data distribution \\
$\zeta$ & the number of parameters of the local model update \\
$R^*$ & the rounds the clustering algorithm needs for finding the cluster centers in XMAM \\
\hline
\end{tabular}
}
\end{center}
\end{table}
\end{document}